
\input jnl
\input eqnorder
\input reforder
\def\sec#1{Sec.\thinspace(#1)}
\def\ffrac#1#2{\textstyle{#1\over#2}\displaystyle}
\preprintno{UCSBTH-93-12}
\preprintno{ISAS-93-75}
\title
Universal Properties of Self-Avoiding Walks from Two-Dimensional
Field Theory
\author
\phantom{xxx}
\author
John Cardy$^{\dag}$
\affil\ucsb
\author
\phantom{xxx}
\author
G. Mussardo
\affil
International School for Advanced Studies
and
Istituto Nazionale di Fisica Nucleare
34013 Trieste, Italy
\abstract\singlespace
We use the recently conjectured exact $S$-matrix of the massive
${\rm O}(n)$ model to derive its form factors and ground state energy.
This information is then used in the limit $n\to0$ to obtain
quantitative results for various universal properties of self-avoiding
chains and loops. In particular, we give the first theoretical prediction of
the amplitude ratio $C/D$ which relates the mean square end-to-end distance
of chains to the mean square radius of gyration
of closed loops. This agrees with the results from
lattice enumeration studies to within their errors, and gives strong
support for the various assumptions which enter into the field
theoretic derivation. In addition, we obtain results for
the scaling function of the structure factor of long loops,
and for various amplitude ratios measuring the shape of self-avoiding
chains. These quantities are all related to moments of correlation
functions which are evaluated as a sum over $m$-particle intermediate
states in the corresponding field theory. We show that in almost all
cases, the restriction to $m\leq2$ gives results which
are accurate to at least one part in $10^3$. This remarkable fact
is traced to a softening of the $m>2$ branch cuts relative to their
behaviour based on phase space arguments alone, a result which follows
from the threshold behaviour of the two-body $S$-matrix, $S(0)=-1$.
Since this is a general property of interacting 2d field theories, it
suggests that similar approximations may well hold for other models.
However, we also study the moments of the area of self-avoiding loops, and show
that, in this case, the 2-particle approximation is not valid.

\vfill
\tenpoint\noindent $^{\dag}$Address after July 1, 1993: Dept.~of Physics,
Theoretical Physics, 1 Keble Road, Oxford OX1 3NP, UK.
\eject
\body
\oneandahalfspace
\head{1. Introduction}

In the past few years, there has been remarkable progress in
obtaining exact results for two-dimensional field
theories\refto{BPZ,ISZ}.
One application of these is to the scaling limit of classical
statistical mechanics systems, close to a second order phase
transition where the correlation length $\xi$ is much larger
than the microscopic cut-off. Yet, besides checking the values of
critical exponents and other universal finite-size scaling
amplitudes which arise from studies of massless, or conformal,
field theories, there has been little in the way of direct
confrontation with statistical mechanics\refto{Lassig}.
This is partly because
the simplest predictions of {\it massive} quantum field theories
are in terms of their particle content and $S$-matrix elements,
as quantum field theories in $1+1$ dimensions.
In order to make contact with
easily observable quantities relevant to statistical mechanics,
it is necessary to go further and derive off-shell behaviour.
More recently, these problems have
become tractable, at least for a class of integrable theories,
with the development of methods for calculating
off-shell form factors\refto{KB,book} and the technique
of the thermodynamic Bethe ansatz\refto{Z1,Z2}.

However, another obstacle is the lack of precise information with
which to compare such predictions. Aside from the Ising
model\refto{Isingb,Ising}
little is known about off-critical correlation functions in
statistical models. Ideally, of course, one should compare with
real experiments, but unfortunately it is notoriously difficult
to obtain reliable data on two-dimensional systems, uncontaminated
by the effects of impurities and finite-size. While, in principle,
it is possible to carry out high-resolution scattering experiments
with X-rays, such studies appear to have gone out of fashion
with experimentalists.

We are therefore left with comparison with the results of numerical
studies. In contrast with real experiments, these are actually easier
to carry out for large systems in low dimensions. While Monte Carlo
studies can study quite large systems, their results are subject to
statistical errors which make comparison with detailed predictions
of the correlation functions difficult. Since we wish to
study such functions unaffected by finite-size effects, exact
diagonalisation of finite transfer matrices is also not useful for
our purpose. This leaves series expansions. The problem for which
the
longest such expansions have been carried out, and therefore the most
accurate information is available, is that of self-avoiding
walks (see \ref{CG} for a review of the current state of knowledge, and
for further references).
Although this is strictly not a problem in critical behaviour,
its
relation to the critical behaviour of the ${\rm O}(n)$ model in the limit
$n\to0$ was shown long ago by de Gennes\refto{DG},
and it is possible to express
all the quantities relevant to the asymptotic behaviour of $N$-step
walks in the limit $N\to\infty$ in terms of correlation
functions and amplitudes.
The correspondence actually relates the fixed fugacity ensemble, where
each step is weighted by a factor $x$, to the $n\to0$ limit of the
${\rm O}(n)$ model, at a temperature simply related to $x$.
One result of this is that the actual correlation length of the ${\rm
O}(n)$ model is not directly measurable through self-avoiding walks.

The field theory of the two-dimensional ${\rm O}(n)$ model has been
studied extensively. While for values of $n>2$ this model
is always in its disordered phase at non-zero temperature, its
analytic continuation to $-2\leq n\leq2$ may be carried out, and
it does then have a finite $T_c$. In fact, the lattice version of this model
at criticality may be mapped {\it via} standard techniques onto the Coulomb
gas model\refto{Nienhuis1,Nienhuis2}, with the result that
the critical exponents previously conjectured\refto{Nijs,CH} and the value
of the central charge $c$ in the corresponding conformal field theory may
be easily read off. The standard methods of conformal field
theory\refto{DF1,Singh} then show that the energy operator corresponds to
a degenerate representation of the Virasoro algebra, labelled by its
position $(1,3)$ in the Kac table. It was shown by A.
Zamolodchikov\refto{Zam1} that a conformal field theory
perturbed by a relevant operator of this type is integrable, in the sense that
an infinite number of the conserved charges in the conformal field
theory survive the perturbation. If the perturbed theory is massive,
which we expect to be the case for $T>T_c$ in the ${\rm O}(n)$ model,
the existence of these conserved charges implies that scattering
in the theory is purely elastic, and that the many-body $S$-matrix
factorises into a product of two-body scatterings\refto{ZZ}. For the
${\rm O}(n)$ model one expects to find a simple particle spectrum:
an $n$-plet of particles states transforming according to the vector
representation of ${\rm O}(n)$, and, since the interaction is repulsive,
no bound states.

On this basis, A. Zamolodchikov\refto{Zpoly} conjectured an exact $S$-matrix
for this theory, based on the principles of analyticity, unitarity,
crossing, and the Yang-Baxter equations. His result represents the
simplest solution to these conditions, but it is not unique, because
of the CDD ambiguity.

Once the $S$-matrix is assumed, however, there is a well-defined
program for investigating the off-shell behaviour of the theory, through the
form factors\refto{KB,book}. For a local field $\phi(r)$, these are defined as
matrix elements $\langle 0|\phi(0)|\beta_1, a_1;\ldots;\beta_m,
a_m\rangle$, where the ket represents an asymptotic $m$-particle state
labelled by the rapidities $\beta_j$ and ${\rm O}(n)$
colour indices $a_j$ of the
particles. The form factors satisfy a set of equations relating them
to the $S$-matrix, and other requirements of analyticity and crossing.
Once again, it is generally assumed that the simplest, or `minimal',
solution of these equations should be chosen. This program has been carried
through in detail only for theories with non-degenerate single-particle
states\refto{YL,CMform,FMS,KM}, or with only simple
symmetries\refto{KB,book,FF1,FF2}. The spectrum
of the ${\rm O}(n)$ model is, however, more complicated, and the two-body
$S$-matrix acts on the direct product of two $n$-dimensional
representations. While, for specific integer values of $n$, it is
clearly possible to decompose this into irreducible amplitudes for which
the $S$-matrix is diagonal, this is not useful for continuation to $n=0$.
We are therefore forced into additional complications in order to include
the ${\rm O}(n)$ structure.

Fortunately, in most cases this may be avoided for all intents and
purposes. The purpose of computing the form factors is to reconstitute
the two-point correlation functions through the unitarity sum
$$
\langle\phi(r)\phi(0)\rangle=\sum_{m=0}^\infty\,\sum_{\rm colours}
\int_{\rm phase\ space}
|\langle0|\phi|\beta_1,a_1;\ldots;\beta_m,a_m\rangle|^2
\,e^{-Mr\sum_{j=1}^m\cosh\beta_j}
\quad.
\eqno(1)
$$
Clearly, for the infrared behaviour (large $r$) only the lowest values
of $m$ are important, while, in the ultraviolet region of small $r$,
in principle all the intermediate states contribute. However, for our
purposes, we shall be interested in moments of the correlation functions
of the form $\int r^p\langle\phi(r)\phi(0)\rangle d^2\!r$. For large
$p$, the large $r$ end of the integration region will be emphasised,
and the truncation to, say, $m\leq2$ should be valid.
Remarkably, however, we find that, this can be true for $p=2$ and
even $p=0$. In the case when $\phi$ is the energy operator (which
couples only to states with $m$ even), the value of the $p=2$ moment
is known exactly from a sum rule derived from the $c$-theorem\refto{cth,CJP}.
The result obtained from the truncation to two-particle states agrees with
this up to about one part in $10^3$. This approximation is less good
for the $p=0$ moment (having an error of about 10\%), but, as will
be discussed below, this moment may be found exactly from the
TBA. When $\phi$ is the ${\rm O}(n)$ vector field itself (corresponding
to the spin of the lattice ${\rm O}(n)$ model), it couples only to states
with odd $m$. Then the truncation to $m=1$ appears to be very accurate
even for the zeroth order moment.

The reason for this remarkable numerical simplification appears to be
that the form factors to the states with higher numbers of particles
have a much softer behaviour close to the $m$-particle threshold
than expected on the basis of phase space alone.
In fact, each form factor contains an explicit factor of
$\prod_{i<j}(\beta_i-\beta_j)$. This may be traced to the threshold
behaviour of the two-body $S$-matrix, $S(0)=-1$. In general, unitarity
implies that $S(\beta)S(-\beta)=1$, so that, in principle, $S(0)$ could
be $\pm1$. However, the upper sign appears to be realised only in a free
boson theory: otherwise, in all interacting theories, the lower sign
holds. Thus, this suppression of the higher particle states would
appear to be a very general feature of two-dimensional theories.

In the case of the ${\rm O}(n)$ model, the restriction to $m\leq2$ makes
the otherwise cumbersome problem of the group theory factors
straightforward. As a result, we are able to obtain accurate values
for the moments of the spin-spin and energy-energy correlation functions
in the $n\to0$ limit, based, of course, on all the assumptions of
minimality for the $S$-matrix and the form factors referred to above.
As will be shown in detail in the subsequent sections, these moments
are simply related, respectively, to moments of the
end-to-end distance of self-avoiding walks, and of the mean square
distance between points of self-avoiding loops.
For these lattice problems, extensive numerical work has been carried
out on the mean square end-to-end distance of $N$-step walks,
$\langle R^2_e\rangle_N\sim CN^{2\nu}$\refto{C}, and the mean square radius
of gyration of loops, $\langle R^2_g\rangle_N\sim DN^{2\nu}$\refto{Privman},
where,
from Coulomb gas arguments, $\nu=\frac34$\refto{Nienhuis1},
and $C$ and $D$ are
amplitudes. By themselves, they are not universal\refto{Privman}
(as will also follow
from our analysis below), but the ratio $C/D$ should be free of all
metric factors and therefore universal.

The only quantity entering the calculation of this ratio which is not
determined to high accuracy by the 2-particle truncation of the form factor
approach is the zeroth moment $U_0$ of the energy-energy correlations. But
this is just proportional to the specific heat, and therefore may be
found in an alternative way by differentiating the extensive part of the
free energy. This may be derived from the thermodynamic Bethe ansatz. The
TBA has been applied to the unitary minimal models with $c=1-6/(k+1)(k+2)$,
with $k$ integer, perturbed by the $(1,3)$ operator, by
Al. Zamolodchikov\refto{Z2}.
Although the ${\rm O}(n)$ model is not identical in its operator content to
the minimal models, the results for the free energy, when perturbed by
the same type of operator, must be the same. Thus we may take over
Zamolodchikov's result, the only generalisation necessary being the
continuation to non-integer $k$, so as to allow the $n\to0$ limit to be
taken. This turns out to be straightforward, and leads to an exact
result for the universal part of the free energy per correlation volume,
which is related to $U_0$.

Piecing together all this information, we find $C/D\approx13.70$.
This is to be compared with the estimates\refto{CG} of $13.69$ for the square
lattice, and $13.72$ for the triangular lattice.
Thus the errors from the 2-particle truncation are less than the
systematic errors of the current data from enumerations.
This agreement is a non-trivial test of the ${\rm O}(n)$ model $S$-matrix
and form factors. (By contrast, the ratio $C/D$ for ordinary random
walks is $12$.)

The mean square radius of gyration of loops is related to the
ratio of the $p=2$ and $p=0$ moments of the energy-energy correlation
function. The generating function for all the moments is proportional
to the mean structure factor of $N$-step loops,
${\cal S}(q)=N^{-1}\sum_{ij}\langle e^{iq\cdot(r_i-r_j)}\rangle$,
which could be measured, for example, by light scattering.
Since the 2-particle approximation gives the zeroth and second moments
to such remarkable accuracy, and its accuracy is supposed to improve
for the higher moments, this gives a very precise way of computing
${\cal S}(q)$, at least for moderate values of its scaling argument
$q^2\langle R^2_g\rangle$. Thus we are able to make the first
accurate calculation of this function. The higher moments are dominated
by the behaviour of the 2-particle cut near threshold. Since, as we
remarked above, this
is softened due to the fact that $S(0)=-1$, these moments behave as a
function of $p$ in a very different way from those of a free theory.
Such a free theory describes random self-intersecting loops. Thus we
are able to picture in a rather direct way how the condition of
self-avoidance influences the statistics of the loops.
On the other hand, the behaviour of $S(q)$ at large values of its
scaling variable $qR$ is proportional to $(qR)^{-1/\nu}$, as expected
for an object of fractal dimension $1/\nu$. The coefficient of this
behaviour may be determined by comparing conformal perturbation theory with
the results of the thermodynamic Bethe ansatz.

Although the truncation to states with $m\leq2$ works remarkably well
for moments of the spin-spin and energy-energy correlations,
this is not true for all correlation functions. For example, the ${\rm O}(n)$
model possesses conserved currents whose integrals generate the
${\rm O}(n)$ symmetry. It turns out that moments of the current-current
correlation function are related to moments of the {\it area}
$\langle a^p\rangle_N$ of $N$-step self-avoiding loops, in the limit
$n\to0$\refto{CG,Miller}. These are expected to scale as
$E^{(p)}N^{p\nu}$\refto{CG,AREA}. Thus we might expect to be
able to compute the amplitudes $E^{(p)}$ in the 2-particle
approximation. The amplitudes of both the first and second moments have
been estimated from enumerations. However, it turns out that
for these values of $p$, the 2-particle approximation is hopeless. This may
be traced to two causes: first, current conservation forces an additional
softening of the 2-particle contribution to the current-current correlation
function near threshold; and second, the fact that the current has unit
scaling dimension means that the ultraviolet behaviour is more singular, and
gives a large contribution to the lower moments which is not adequately
represented by the 2-particle contribution.

While the form factor approach has, so far, been applied to the
computation of two-point functions, it is not restricted to this case.
Several amplitudes governing the statistics of open self-avoiding walks,
for example their mean square radius of gyration, are
related to moments of higher-point correlation functions. We show how
in principle this calculation may be carried out, and that, in the
simplest approximation, we get results which agree with those of
numerical lattice calculations to within a few percent.

The lay-out of this paper is as follows. In the next section, we
make precise the connection between the self-avoiding walk problem and
the ${\rm O}(n)$ model, and in particular relate the various amplitudes and
scaling functions whose values we are trying to compute to moments
of appropriate correlation functions in the field theory. In \sec{3}
we recall A. Zamolodchikov's arguments\refto{Zpoly} for the $S$-matrix of this
theory, and derive the 2-particle form factors. We show how this
leads to an accurate estimate for the second moment of the energy-energy
correlation function. In the next section we derive the zeroth moment
of this correlation function by way of the thermodynamic Bethe ansatz,
generalizing the arguments of Al. Zamolodchikov\refto{Z2} so that the limit
$n\to0$ may be taken. This result, combined with those of the previous
section, then leads to our estimate for $C/D$.

In \sec{5} we present the calculation of the structure factor and
comment on its features, and in \sec{6} we discuss the current-current
correlation function and show why the 2-particle approximation fails for
its lower moments. In \sec{7} we consider the radius of gyration of
open chains, which involves the calculation of a three-point correlation
function, and, finally, we summarise our conclusions and discuss
possible further extensions of our results.

\head{2. The ${\rm O}(n)$ model and self-avoiding walks}

In this section we summarize the well-known relation\refto{DG} between these
two problems in the limit $n\to0$, in order to make precise the
quantities we shall need later.

Consider first the set of $N$-step self-avoiding walks on a given
lattice. It is convenient to consider lattices of co-ordination number
three ({\it e.g.} a honeycomb lattice in two dimensions), but, later on,
an appeal to universality will imply that in the critical region the
results are universal. In the same spirit let us consider a specific
lattice model with ${\rm O}(n)$ symmetry, in which the degrees of freedom
are spins $s_a(r)$ at each site $r$, labelled by a `colour' index
$a$ which takes values from 1 to $n$. The spins are normalised so that
\def\tr{{\rm Tr}\,}
$\tr{s_as_b}=\delta_{ab}$, and the partition function is
$$
Z=\tr\prod_{\rm nn}\left(1+x\sum_as_a(r_i)s_a(r_j)\right)
\quad,\eqno(a)
$$
where the product is over all nearest neighbour pairs of sites
$(r_i,r_j)$. Once again, we expect, on the grounds of universality,
that the particular form of the hamiltonian\break
$H_1=-\sum_{\rm nn}\ln\left(1+x\sum_as_a(r_i)s_a(r_j)\right)$
implied by \(a) should
not be important in the critical region. In particular, we could
consider the more standard hamiltonian\break $H_2=-x\sum_{\rm nn}
\sum_as_a(r_i)s_a(r_j)$.

The continuum limit of either of these models close their respective
critical points is assumed to be described by an ${\rm O}(n)$-invariant
field theory, and, at the critical point, by a conformal field theory
whose central charge $c(n)= 1-6/(k+1)(k+2)$, where $n=2\cos(\pi/(k+1))$ with
$k\geq 0$\refto{DF1,Singh}. We note, in particular, that $c$ vanishes when
$n=0$, as it should, since it measures the finite-size correction to the free
energy which would vanish in this case. Its derivative $dc/dn$ at $n=0$
will be important: this takes the value $5/3\pi$. The scaling dimension of the
energy operator is $x_e=2k/(k+2)$ and equals $2/3$ for $n=0$. It is
related to the exponent $\nu$ characterising the divergence of the
correlation length by $\nu=1/(2-x_e)=3/4$. This also fixes the critical
index $\alpha=2-2\nu$ of the singular part of the free energy. The other
exponents which will enter are $\gamma=43/32$ which gives the divergent
behaviour $(x_c-x)^{-\gamma}$ of the susceptibility in the limit $n=0$, and
the magnetic scaling index $\eta=5/24$. They are related by the scaling
equation $\gamma=\nu(2-\eta)$.

On expanding the product in \(a), it may be written as a sum over
self-avoiding loops:
$$
Z=\sum_{{\rm loop}\atop{\rm configurations}}x^{\rm number\ of\ links}
\,n^{\rm number\ of\ loops}
\quad.\eqno(b)
$$
As $n\to0$, only configurations with a single loop survive in the
$O(n)$ term. Thus, if $p_N$ is defined as the number of $N$-step
self-avoiding
loops per lattice site, then
$$
{\cal N}_s\sum_Np_Nx^N=\lim_{n\to0}n^{-1}\ln Z
\quad,\eqno(c)
$$
where ${\cal N}_s$ is the total number of lattice sites. Note that
the right hand side is proportional to the free energy, so is also
proportional to the total area $\cal A$.
The free energy of the
${\rm O}(n)$ model is believed to be an analytic function of its
temperature-like variable $x$ for sufficiently small $x$, and to
have a singularity of the form $(x_c-x)^{2-\alpha}$ at some finite,
lattice-dependent $x_c$. Since the coefficients of the series on the
left hand side are all non-negative, the singularities of the free energy
on the circle $|x|=x_c$ will determine the behaviour of $p_N$ at large $N$.
Hyperscaling implies that, in two dimensions,
${\cal A}^{-1}\ln Z\sim nU\xi^{-2}$, where $\xi$ is the correlation
length and $U$ (which we shall calculate exactly in \sec{4}) is
universal. The correlation length itself has the critical behaviour
$$
\xi\sim\xi_0(1-x/x_c)^{-\nu}
\quad,\eqno()
$$
where $\nu=3/4$ but the metric factor $\xi_0$ is non-universal.
Putting all these factors together we see that
$$\sum_Np_Nx^N\sim a_0\xi_0^{-2}U\,(1-x/x_c)^{2\nu}
\quad,\eqno(d)
$$
as $x\to x_c$,
where $a_0={\cal A}/{\cal N}_s$ is the area per site. This implies
that, as $N\to\infty$, $p_N\sim BN^{-2\nu-1}\mu^N$, where $\mu=x_c^{-1}$
and the the amplitude $B$ is given by
$$
B=\sigma a_0\xi_0^{-2}{U\over\Gamma(-2\nu)}
\quad.\eqno(e)
$$
For certain lattices $p_N$ actually vanishes by symmetry except when
$N$ is divisible by an integer $\sigma$. This implies that the
left hand side of \(d) is actually a series in $x^\sigma$, and therefore
has $\sigma$ equivalent singularities on its circle of convergence, thus
accounting for the lattice-dependent factor of $\sigma$ on the right
hand side of \(e). For example, for the square lattice $\sigma=2$.

Next, we consider the energy-energy correlation function. Close to the
critical point, we write $H_2=H_2^c+ (x_c-x)\sum_{{\rm bonds\ } r}
{\cal E}^{\rm lat}(r)$, where ${\cal E}^{\rm lat}(r)=\sum_as_a(r_<)s_a(r_>)$
is the lattice energy located on the link $r$ which connects the sites
$r_<$ and $r_>$.
The correlation function $\langle {\cal E}^{\rm lat}(r_1)\,
{\cal E}^{\rm lat}(r_2)\rangle$
then receives contributions from all self-avoiding loops which contain
the links $r_1$ and $r_2$. If the number of such loops with $N$ steps is
$f_N(r_1,r_2)$, then
$$
n\sum_Nf_N(r_1,r_2)x^{N-2}=\langle {\cal E}^{\rm lat}(r_1)\,\,{\cal E}^{\rm
lat}(r_2)\rangle
\quad.\eqno(f)
$$
In particular, if we sum over all $r_1\not=r_2$ on the left hand side,
we sum over all such $N$-step loops, and obtain
$$
n{\cal N}_s\sum_NN(N-1)p_Nx^{N-2}=\sum_{r_1\not=r_2}
\langle {\cal E}^{\rm lat}(r_1) \,\,{\cal E}^{\rm lat}(r_2)\rangle
\quad,\eqno(g)
$$
which is just the second derivative of \(c) with respect to $x$. The
right hand side is just the specific heat of the ${\rm O}(n)$ model.

In the same way, if we multiply by $(r_1-r_2)^2$ before summing,
we get the generating function for $p_N$ weighted by
$$
\sum_{r_1,r_2}(r_1-r_2)^2=\sum_{r_1,r_2}\bigg((r_1-\bar r)-
(r_2-\bar r)\bigg)^2
=2N^2R^2_g
\quad,\eqno(h)
$$
so that
$$
2n{\cal N}_s\sum_NN^2p_N\langle R^2_g\rangle_Nx^{N-2}=\sum_{r_1,r_2}
(r_1-r_2)^2\langle {\cal E}^{\rm lat}(r_1) \,\,{\cal E}^{\rm lat}(r_2)\rangle
\quad.\eqno(i)
$$
Similarly, the generating function for the structure factor
${\cal S}_N(q)
=N^{-1}\sum_{r_1,r_2}\langle e^{iq\cdot(r_1-r_2)}\rangle$
is given by
$$
n{\cal N}_s\sum_NNp_N{\cal S}_N(q)x^{N-2}=\sum_{r_1,r_2}
e^{iq\cdot(r_1-r_2)}\langle {\cal E}^{\rm lat}(r_1) \,\,
{\cal E}^{\rm lat}(r_2)\rangle
\quad.\eqno(j)
$$
In the scaling region, the dominant contribution to the sum on the
right hand side comes from $|r_1-r_2|$ of the order of the correlation
length, and we may therefore use a continuum field theory description.
The lattice energy ${\cal E}^{\rm lat}$ is replaced by the continuum energy
density ${\cal E}(r)$ in such a way that they enter into the hamiltonian
(action) in the same way:
$$
\sum_r {\cal E}^{\rm lat}(r)\to \int {\cal E}(r)\,d^2\!r
\quad.\eqno(k)
$$
However, in the field theory, the energy operator is special in that
it represents the perturbation $(x_c-x)\int {\cal E}(r)\,d^2\!r$ of the
critical
theory. It is therefore proportional to the trace $\Theta(r)=T^\mu_\mu$ of the
stress tensor. Explicitly
$$
\Theta(r)=-2\pi\nu^{-1}(x_c-x)\,{\cal E}(r)
\quad,\eqno(l)
$$
where the conventional factor of $2\pi$ has been included in the
definition of the stress tensor. Since the normalisation of the stress tensor
is fixed, its correlation functions are completely universal. In particular
$$
\langle\Theta(r)\Theta(0)\rangle=n\xi^{-4}\Phi_1(r/\xi)
\quad,\eqno(m)
$$
where $\Phi_1$ is a universal scaling function.

{}From \(i) we then see that, in terms of this scaling function,
$$
\eqalignno{
2n\sum_NN^2p_N\langle R^2\rangle_Nx^{N-2}
&=a_0(\nu/2\pi)^2(x_c-x)^{-2}\int r^2\langle\Theta(r)\Theta(0)\rangle d^2r
&(n)\cr
&=na_0(\nu/2\pi)^2U_2(x_c-x)^{-2}&(o)\cr}
$$
where $U_2=\int\rho^2\Phi_1(\rho)d^2\rho$ is universal, so that
$$
2Np_N\langle R^2_g\rangle_N\sim \sigma a_0(\nu/2\pi)^2U_2\mu^N
\quad.\eqno(p)
$$
The right hand side of \(n) is in fact given by a sum rule which is a
consequence of Zamolodchikov's $c$-theorem\refto{cth}:
$$
c(n)=(3/4\pi)\int r^2\langle\Theta(r)\Theta(0)\rangle d^2r
\quad,\eqno(q)
$$
where $c(n)$ is the central charge of the ${\rm O}(n)$ model.
In terms of the amplitude $D$ in the relation $\langle R^2_g\rangle
\sim DN^{2\nu}$, this implies the relation
$$
BD={5\over 32\pi^2}\sigma a_0
\quad,\eqno(r)
$$
which was derived in \refto{CJP}, and generalised in \refto{CG}. Alternatively
we see that the $c$-theorem sum rule gives an exact value for $U_2$:
$$
U_2=\int\rho^2\Phi_1(\rho)d^2\rho=\frac{20}9
\quad.\eqno(s)
$$

{}From \(g), the amplitude $B$ may itself be written in terms
of the zeroth moment of $\Phi_1$, since
$$
\sum_NN^2p_Nx^{N-2}=a_0(\nu/2\pi)^2(x_c-x)^{-2}
\int\langle\Theta(r)\Theta(0)\rangle d^2r=
a_0(\nu/2\pi)^2(x_c-x)^{-2}\xi^{-2}U_0
$$
where $U_0=\int\Phi_1(\rho)d^2\rho$. Hence we obtain the alternate
result for the amplitude $B$:
$$
B=\sigma a_0\xi_0^{-2}(\nu/2\pi)^2{U_0\over\Gamma(2-2\nu)}
\quad.\eqno(t)
$$
Comparing with \(e), this gives the further sum rule
$$
U_0=\int\Phi_1(\rho)d^2\rho=(2\pi/\nu)^2(-2\nu)(1-2\nu)\,U
\quad.\eqno(u)
$$
We stress that this is merely a consequence of the fluctuation sum
in \(g) being proportional to the specific heat. As we shall see in \sec{4},
the amplitude $U$ may be obtained exactly. The above two sum rules for
$U_0$ and $U_2$ then give an estimate of the errors of the
two-particle approximation to $\langle\Theta(r)\Theta(0)\rangle$.

The amplitude $D$ defined above
for the radius of gyration may also be written,
from \(e,p,u)
in the more intuitive form
$$
D=\xi_0^2\,{\Gamma(\alpha)\over2\Gamma(\alpha+2\nu)}{U_2\over U_0}
\quad,\eqno(v)
$$
As expected, the mean square radius of gyration
is proportional to the ratio of the second to the zeroth
moments of the energy-energy correlation function.

Finally, the structure factor is related to the Fourier transform
of $\Phi_1$ by
$$
\sum_NNp_N{\cal S}_N(q)x^{N-2}=a_0(\nu/2\pi)^2(x_c-x)^2\xi^{-2}
\int e^{iq\cdot\rho\xi}\,\,\Phi_1(\rho)d^2\rho
\quad.\eqno(w)
$$
Disentangling the large $N$ dependence from this equation is more
complicated, and will discussed in \sec{5}.

Next, we consider the case of linear self-avoiding walks. Let $c_N(r)$
be the number of such walks with $N$ steps from the origin to the site
$r$. The generating function for this is just the spin-spin correlation
function of the ${\rm O}(n)$ model as $n\to0$:
$$
\sum_Nc_N(r)x^N=\lim_{n\to0}\langle s_1(r)s_1(0)\rangle
\quad.\eqno(x)
$$
In the scaling region, we expect this correlation function to scale
as
$$
\langle s_1(r)s_1(0)\rangle\sim b\xi^{-\eta}\Phi_2(r/\xi)
\quad,\eqno(y)
$$
where $\Phi_2$ is universal, but the metric factor $b$ is not.
Note that the magnetisation is different from the energy operator in that
it has no absolute normalisation. From \(x) we may easily compute the
generating function for the number $c_N$ of all $N$-step walks:
$$
\sum_Nc_Nx^N\,=\,b\,\xi_0^{2-\eta}(1-x/x_c)^{-\gamma}\,V_0
\quad,\eqno(z)
$$
and for this number weighted by their squared end-to-end distance
$R^2_e$:
$$
\sum_Nc_N\langle
R^2_e\rangle_Nx^N=b\,\xi_0^{4-\eta}(1-x/x_c)^{-\gamma-2\nu}\,V_2
\quad,\eqno(aa)
$$
where $V_0$ and $V_2$ are the zeroth and second moments, respectively,
of $\Phi_2$.
We then find, for the amplitude $C$ in the asymptotic behaviour
$\langle R^2_e\rangle\sim CN^{2\nu}$, the result
$$
C=\xi_0^2\,{\Gamma(\gamma)\over\Gamma(\gamma+2\nu)}{V_2\over V_0}
\quad.\eqno(ab)
$$
Note the close similarity with \(v). (The extra factor of $2$ in the
denominator in \(v) is due to the factor of $2$ appearing in \(h).)

Neither of the amplitudes $C$ or $D$ is universal, because of the
metric factor $\xi_0^2$. However, this cancels in their ratio,
and we find
$$
C/D={2\Gamma(\gamma)\Gamma(\alpha+2\nu)\over\Gamma(\gamma+2\nu)
\Gamma(\alpha)}{V_2\,U_0\over V_0\,U_2}
\quad.\eqno(ac)
$$

\head{3. $S$-matrix and Form Factors of the Thermal Perturbation of the
${\rm O}(n)$ model}
\subhead{3.1 Scattering theory}
Since in the scaling limit the energy operator of the ${\rm O}(n)$
model on the
lattice corresponds to the primary operator $\varphi_{1,3}$ of the conformal
model, the thermal perturbation of the critical point action is described by
$$
A=A_c+(x_c-x)\,\int\,\varphi_{1,3}(x)\,d^2x
\quad.
\eqno(action)
$$
As with any $\varphi_{1,3}$ deformation of the minimal conformal models, the
QFT
defined by the action \(action) presents a number of higher
integrals of motion which guarantee its integrability\refto{Zam1}. For $x<x_c$
(the only case that we will investigate) the model develops a finite
correlation length and the associate scaling QFT is mainly characterised by
the factorised $S$-matrix of its massive excitations. The scattering theory
has been proposed by Zamolodchikov\refto{Zpoly} and its main
features may be summarised as follows\footnote{$^1$}{It is worth mentioning
that the same $S$-matrix, although with a different interpretation, has been
also derived by Smirnov\refto{Smirnovpoly} using a quantum group
reduction of the scattering theory of the Sine-Gordon model.}.

First of all, on the basis of the form of the partition function \(c),
Zamolodchikov argued that it is possible to interpret the loops as
trajectories of a set of $n$ particles $A_i(\beta)$ (with mass M) that belong
to the vector representation of ${\rm O}(n)$.
Hence the scattering matrix for the process
$|A_{i_1}(\beta_1) A_{i_2}(\beta_2)\rangle \rightarrow |A_{j_1}(\beta_1)
A_{j_2}(\beta_2)\rangle$ may be written as
$$
S_{i_1i_2}^{j_1j_2}(\beta_{12}) =
S_0(\beta_{12})\, \delta_{i_1}^{j_1} \delta_{i_2}^{j_2} +
S_1(\beta_{12})\, \delta_{i_1}^{j_2} \delta_{i_2}^{j_1} +
S_2(\beta_{12}) \,\delta_{i_1i_2} \delta^{j_1j_2}
\quad,
\eqno(smatrix)
$$
where $\beta_{12}=\beta_1-\beta_2$.
$S_0$ is the amplitude for the transmission process whereas $S_1$ and
$S_2$ are respectively the amplitudes for the reflection and annihilation
processes. Equivalently, we may decompose the scattering matrix
into channels of definite isospin: for the symmetric traceless,
antisymmetric and isosinglet channels we have respectively
$$
\eqalign{
& S_S(\beta)\,=\, S_0(\beta)\,+\,S_1(\beta) \cr
& S_{A}(\beta)\,=\, S_0(\beta)\,-\,S_1(\beta) \cr
& S_{I}(\beta)\,=\,S_0(\beta)\,+\,S_1(\beta)\,+\,n\,S_2(\beta)\quad.\cr}
\eqno(colchannels)
$$
The amplitudes $S_1$ and $S_2$ are linked to each other by
the crossing symmetry relation
$$
S_1(\beta) = S_2(i\pi-\beta)
\quad,
\eqno(crossing)
$$
whereas $S_0$ is a crossing symmetric function.

In order to take into account the property that the loops entering \(c) are
non-intersecting paths, Zamolodchikov suggested imposing the condition
$$
S_0(\beta)=0
\quad.
\eqno(selfavoiding)
$$
Using the Yang-Baxter equations and the unitarity condition,
the final form of the minimal $S$-matrix is then given by
$$
\eqalign{
& S_1(\beta) = - \sinh\left(\frac{i\pi-\beta}{k+1}\right)R(\beta)\cr
& S_2(\beta) = - \sinh\left(\frac{\beta}{k+1}\right)R(\beta)\quad,\cr}
\eqno(starr)
$$
where
$$
R(\beta)=\frac{1}{\sinh\left(\frac{i\pi-\beta}{k+1}\right)}
\exp\left[i\int_0^{\infty}\frac{dx}{x}\frac{\sinh\frac{\pi k x}{2}
\sin x\beta}{\sinh\frac{\pi (k+1) x}{2}\cosh\frac{\pi x}{2}}
\right]
\quad.
\eqno(Rbeta)
$$
Its analytic structure may be read off from its infinite product representation
$$
\eqalign{
R(\beta) =&  \frac{1}{\sin\pi\left(\frac{1}{k+1}-\frac{\beta}{i\pi
(k+1)}\right)} \frac{\Gamma\left(1-\frac{\beta}{i\pi (k+1)}\right)}
{\Gamma\left(1+\frac{\beta}{i\pi (k+1)}\right)}
\prod_{l=1}^{\infty}
\frac{\Gamma\left(\frac{2l}{k+1}-\frac{\beta}{i\pi (k+1)}\right)}
{\Gamma\left(\frac{2l}{k+1}+\frac{\beta}{i\pi (k+1)}\right)}
\cr
& \times\,
\frac{\Gamma\left(1+\frac{2l}{k+1}-\frac{\beta}{i\pi (k+1)}\right)
\Gamma\left(\frac{2l-1}{k+1}+\frac{\beta}{i\pi (k+1)}\right)
\Gamma\left(1+\frac{2l-1}{k+1}+\frac{\beta}{i\pi (k+1)}\right)}
{\Gamma\left(1+\frac{2l}{k+1}+\frac{\beta}{i\pi (k+1)}\right)
\Gamma\left(\frac{2l-1}{k+1}-\frac{\beta}{i\pi (k+1)}\right)
\Gamma\left(1+\frac{2l-1}{k+1}-\frac{\beta}{i\pi (k+1)}\right)}
\quad.\cr}
\eqno(Rinfinity)
$$
Since there are no poles in the physical sheet $0\leq {\rm Im}\beta\leq \pi$,
no bound states appear. Hence the whole particle content of the model only
consists in the $n$ degenerate states of the vector representation of
${\rm O}(n)$.
For $n=1$ the $S$-matrix of the isosinglet channel correctly coincides with
the $S$-matrix of the thermal perturbation of the Ising model
$$
S_1(\beta)+S_2(\beta)=-1
\quad,
$$
whereas in the limit $n\rightarrow 0$, the final form of the scattering
amplitudes is given by
$$
\eqalign{
S_1(\beta) = & -G(\beta) \cr
S_2(\beta) = & -i\tanh\left(\frac{\beta}{2}\right)G(\beta)\quad, \cr}
\eqno(spolymer)
$$
where
$$
G(\beta)=
\exp\left[i\int_0^{\infty}
\frac{dx}{x}\frac{\sinh\frac{\pi x}{2}
\sin x\beta}{\sinh\pi x\cosh\frac{\pi x}{2}}
\right]
\quad.
\eqno(Gpoly)
$$
Notice that the $S$-matrix of the ${\rm O}(n)$ model possesses nontrivial
asymptotic behaviour for $\beta\rightarrow\pm \infty$
$$
\eqalign{
S_1(\beta) \rightarrow & \,e^{\pm i\pi\Delta} \cr
S_2(\beta) \rightarrow & \mp\,e^{\pm i\pi(\Delta+\frac{1}{k+1})}\quad,\cr}
\eqno(asymdelta)
$$
where
$$
\Delta = \frac{3 k +2 }{2 (k+1)}
\quad.$$
According to\refto{FF2}, this implies that the particle operators
of the ${\rm O}(n)$ model satisfy a generalised statistics.
\subhead{3.2 Form Factors}
The previous discussion of the scattering theory of the thermal deformation
of the ${\rm O}(n)$ model gives rise to a series of questions related to
the minimality assumptions made in its derivation or to the non-integer values
assumed by the variable $n$. In addition the final scattering amplitudes
are not directly related to the underlying microscopic formulation of the
model, making it apparently problematic to judge their validity. To answer
such questions, it is useful to recall that for integrable models the
knowledge of the $S$-matrix is a powerful starting point to compute their
correlation functions through the form factor bootstrap
approach\refto{KB,book}. Hence we may take the point of view
of considering the ${\rm O}(n)$ scattering theory discussed above as
simply the basic tool to implement the program of reconstruction of correlation
function for this model, the whole justification
of the approach being in the final
comparison with quantities extracted by other methods.

In the form factor bootstrap approach the correlation functions are computed
by exploiting their spectral representations, that is, their expression as an
infinite series over multi-particle intermediate states. For instance,
the two-point function of an isoscalar operator $\Phi(x)$ in real Euclidean
space is given by
$$
\eqalign{
\langle\Phi(x)\Phi(0)\rangle= &
\sum_{n=0}^{\infty}
\int \frac{d\beta_1\ldots d\beta_n}{n! (2\pi)^n}
<0|\Phi(x)|\beta_1,a_1;..\beta_n,a_n>
<\beta_1,a_1..\beta_n,a_n|\Phi(0)|0>
\cr
& =\sum_{n=0}^{\infty}
\int \frac{d\beta_1\ldots d\beta_n}{n! (2\pi)^n}
\mid F^{\Phi}_{a_1,\dots,a_n}(\beta_1\ldots \beta_n)\mid^2
\exp \left(-Mr\sum_{i=1}^n\cosh\beta_i \right)\quad,\cr}
\eqno(spectral)
$$
where a sum on the colour indices is implied. In
\(spectral) $r$ denotes the radial distance, i.e. $r=\sqrt{x_0^2 + x_1^2}$ and
$$
F^{\Phi}(\beta_1,\ldots,\beta_n)_{a_1,a_2,\ldots,a_n} \,\equiv\,
<0\mid\Phi(0)\mid \beta_1,a_1;\ldots;\beta_n,a_n>
\eqno(Formfactors)
$$
are the so-called {\it form factors} (FF). The normalization of the
asymptotic states is fixed as
$$
\langle \beta_1,a_1\mid\beta_2,a_2\rangle\,=\,2\pi\,\delta_{a_1,a_2}\,
\delta(\beta_1-\beta_2)
\quad.
\eqno(normalization)
$$
Since the spectral representations are based only on the completeness of the
asymptotic states, they are general expressions for any QFT. However, for
integrable models, they become quite effective because the exact computation
of the form factors reduces to finding a solution of a finite set of
functional equations. In fact they satisfy the Watson equations
\refto{KB,book}
$$
\eqalign{
F^{\Phi}(\beta_1...\beta_{i+1},\beta_i,...
\beta_n)_{a_1..a_{i+1}',a_i'..a_n} = &
F^{\Phi}(\beta_1..
\beta_i,\beta_{i+1}..\beta_n)_{a_1..a_i,a_{i+1}..a_n}
\, S^{a_i,a_{i+1}}_{a_i',a_{i+1}'}
(\beta_i-\beta_{i+1}) \cr
F^{\Phi}(\beta_1,\ldots,\beta_n+2\pi i)_{a_1,\ldots,a_n} = &
F^{\Phi}(\beta_n,\ldots,\beta_{n-1})_{a_n,a_1,\ldots,a_{n-1}}
\quad.\cr}
\eqno(functional)
$$
The first of \(functional) show that the monodromy properties of the FF are
determined by the two-body $S$-matrix of the model. On the other hand the
second
equation states that an analytic continuation in the
variables $\beta_i$ simply induces a reordering in the set of the asymptotic
particles. Notice that this system of equations does not uniquely fixed
the solution since the multiplication of
$F^{\Phi}(\beta_1,\ldots,\beta_n)_{a_1,\ldots,a_n}$ by a symmetric,
$2\pi i$ periodic function leaves \(functional) untouched.
For a specific operator this ambiguity may be generally solved by the
knowledge of the asymptotic behaviour of its form factors and their analytic
structure. Let us briefly discuss the two aspects separately.

The asymptotic behaviour of a FF under a simultaneous shift of all rapidity
variables is simply dictated by relativistic invariance
$$
F^{\Phi} (\beta_1+\Lambda,\beta_2+\Lambda,\ldots,
\beta_n+\Lambda)_{a_1,\ldots,a_n} \,=\,
e^{s\Lambda}\,F^{\Phi} (\beta_1,\beta_2,\ldots,\beta_n)_{a_1,\ldots,a_n}
\quad,
\eqno(asymp2)
$$
where $s$ is the spin of the operator $\Phi$. Notice that for scalar operators
the FF depend only on the differences $\beta_{ij}=\beta_i-\beta_j$.
Secondly, restricting our attention only to those operators which have
two-point functions with a power-law ultraviolet behaviour,
we have to require that their FF behave asymptotically no worse than
$\exp(k \beta_i)$ for $\beta_i \rightarrow \infty$, where $k$ is a constant
independent of $i$. If a perturbative formulation of the
theory is available, the constant $k$ may be fixed by matching the
asymptotic behaviour of the FF with the asymptotic behaviour of the relevant
Feynman diagrams which contribute to the amplitude.

Concerning the analytic nature of the form factors, their pole
structure gives rise to a set of recursive equations which relate
the $l$-particle FF ($l>2$). Since in the ${\rm O}(n)$ model there
are no bound states, the only singularities which appear in the FF are those
associated with the annihilation poles $\beta_a=\beta_b +i\pi$ with
residues\refto{book}
$$
\eqalign{
& 2\pi\, i\,\, {\rm res} F^{\Phi}(\beta + i \pi, \beta,\beta_1,\ldots,
\beta_l)_{a_0,\bar{a}_0,a_1,\ldots,a_l} = \cr
&\left(
\delta_{a_1}^{a_1'} \delta_{a_2}^{a_2'} \ldots \delta_{a_l}^{a_l'}
- S_{a_1,\ldots a_l}^{a_1',\ldots a_l'}(\beta_1,\ldots,\beta_l\mid\beta)
\right) F^{\Phi}(\beta_1,\ldots,\beta_l)_{a_1',\ldots,a_l'} \cr}
\eqno(recursive)
$$
where
$$
S_{a_1,\ldots a_l}^{a_1',\ldots a_l'}(\beta_1,\ldots,\beta_l\mid\beta) =
S^{a_1',\tau_1}_{a_1,a_0}(\beta_1-\beta)
S^{a_2',\tau_2}_{\tau_1,a_2}(\beta_2-\beta)\ldots
S^{a_l',a_0}_{a_l,\tau_{l-1}}(\beta_l-\beta)
\quad.
\eqno(transfer)
$$
These equations induce a recursive structure in the space of the
FF relating matrix elements with $l+2$ and $l$ particles.

After this short discussion on the general properties of the FF, let us
turn our attention to the ${\rm O}(n)$ models. In order to analyse the
correlation
functions in the high-temperature phase of the model, it becomes
convenient to define initially a two-particle form factor $F_{\rm min}(\beta)$
that does not depend on colour indices, does not contain poles on the physical
sheet and has the mildest behaviour for large values of $\beta$. The
first requirement implies that the monodromy property of such FF is
induced by the $S$-matrix of the isosinglet channel $S_I(\beta)$ i.e.
$$
\eqalign{
& F_{\rm min}(\beta)\, = \,[S_1(\beta)+n\,S_2(\beta)]\,F_{\rm min}(-\beta) \cr
& F_{\rm min}(i\pi-\beta)\, = \,F_{\rm min}(i\pi+\beta)\quad. \cr}
\eqno(watsoniso)
$$
The solution of these equations which satisfies the above requirements is
given by
$$
F_{\rm min}(\beta)\,=\,\sinh\frac{\beta}{2} \,f_k(\beta) \,
\exp\left[-\int_0^{\infty}
\frac{dx}{x}\frac{\sinh\frac{\pi k x}{2}\sin^2\frac{\hat\beta x}{2}}
{\sinh\pi x \sinh\frac{\pi(k+1) x}{2}\,\cosh\frac{\pi x}{2}}\right]
\quad,
\eqno(fminn)
$$
where $\hat \beta =i\pi-\beta$ and
$$
f_k(\beta) \,= \,{\sin(\hat\beta/2)\over\sin\big(\hat\beta/(k+1)\big)}
\quad.\eqno(fk)
$$
Note that $f_1(\beta)=1$.
In the self-avoiding case ($n\rightarrow 0$) the monodromy property
of $F_{\rm min}(\beta)$ is dictated only by $S_1(\beta)$ of \(spolymer) and
its expression reduces to
$$
F_{\rm min}(\beta)\,=\,\sinh\frac{\beta}{2}\,\exp\left[-\int_0^{\infty}
\frac{dx}{x}\frac{\sinh\frac{\pi x}{2}\sin^2\frac{1}{2}(i\pi-\beta)x}
{\sinh^2\pi x \,\cosh\frac{\pi x}{2}}\right]
\quad.\eqno(fminpoly)
$$
Useful properties of this function are discussed in the Appendix.

Notice that at the threshold ($\beta\rightarrow 0$) $F_{\rm min}(\beta)$ goes
linearly to zero since $S_I(0)=-1$. The vanishing of the
$F_{\rm min}(\beta)$ at the threshold has far-reaching consequences because
it induces a suppression of the higher particle contributions in the spectral
representation of the correlation functions. In order to see this, let us
consider the general parametrisation of a FF of an operator $\Phi^{a,\ldots,m}$
with colour indices $a,\dots,m$
$$
<0\mid\Phi^{a,\ldots,m}(0)\mid \beta_1,a_1;\ldots;\beta_l,a_l> =
{\cal F}^{a,\ldots,m}_{a_1,\ldots,a_n}(\beta_1,\ldots,\beta_l) \,
\prod_{i<j} \frac{F_{\rm min}(\beta_{ij})}{\cosh\frac{\beta_{ij}}{2}}
\quad.
\eqno(parametrization)
$$
The denominator in \(parametrization) takes into account the
pole structure of these matrix elements whereas the function
${\cal F}^{l,m,\ldots}_{a_1,\ldots,a_n}(\beta_1,\ldots,b_n)$
carries the colour structure of this matrix element. The important
point is that this function has neither poles or zeros in the physical sheet
and satisfies a simplified system of monodromy equations which arise by
substituting \(parametrization) into \(functional).

Let us consider now the contribution of the four-particle intermediate
states to the 2-point functions. If the corresponding form factor was
constant, this contribution would be of the form
$$
\int\prod_1^4d\beta_ie^{-Mr\sum_1^4\cosh\beta_i}
\sim {e^{-4Mr}\over r^2}
\quad,
$$
which corresponds in momentum space to a branch point at $q^2=-16M^2$ of the
form $\ln(q^2+16M^2)$. But since $F_{\rm min}$ vanishes at the threshold,
neglecting other terms, we have an additional factor
$\prod_{i<j}\beta_{ij}^2$ in the integral. This leads to a behaviour
${e^{-4Mr}\over r^8}$, so that the branch cut gets softened to
a function of the form $(q^2+16M^2)^3\ln(q^2+16M^2)$. This effect actually
gets stronger as we consider higher numbers of particles:
the $m$-particle branch cut would be, on the grounds of phase space alone,
of the form $(q^2+m^2M^2)^{m/4-1}$, but it actually gets softened to
$(q^2+m^2M^2)^{m^2/4-1}$.

As it is evident from the above discussion, this suppression of the
higher particle states is not peculiar to the ${\rm O}(n)$ model but on the
contrary we expect it to be a completely general feature for all interacting
theories which have $S(0)=-1$. From a pragmatic point of view, this observation
is crucial to the success of the form factor bootstrap approach to the
computation of correlation functions. In fact, although for integrable models
the exact computation of the FF may be achieved with little effort on the
basis of our previous considerations, one may wonder about their final
usefulness. After all, to compute the correlation functions we have to sum
over the infinite multiparticle FF, and these series are in general hard to
analyse. Of course, for large values of $Mr$ only the lowest term
will dominate the sum, the higher particle contributions being
exponentially small. For small values of $Mr$ however the correlators
possess singularities and, on these scales, all numbers of particles contribute
in principle to the sum. But if there is reason to believe that the
higher particle terms are anyway suppressed, then the lowest contribution
of the spectral representation becomes quite effective in approximating the
correlation functions even in these ultraviolet regions. The validity of
such approximation is of course linked to a sufficiently mild
divergence of the correlation functions in the short distance scales.

A first check of this phenomenon of suppression of higher FF contributions
is given by looking at the second moment of the scaling function $\Phi_1(\rho)$
$$
U_2\,=\,\int \rho^2\,\Phi_1(\rho) \,d^2\rho
\quad.
\eqno(U2)
$$
The extra power in $\rho^2$ helps to smooth the ultraviolet behaviour of the
correlation function and we may expect this integral to be almost saturated
by the two-particle contribution. For the trace of the stress-energy tensor
the two-particle form factor we take\footnote{$^1$}{We assume the mildest
behaviour at infinity of this matrix element to fix it uniquely. Its
normalization is easily determined by using the definition of the
hamiltonian operator
$
H\,=\,\frac{1}{2\pi} \int_{-\infty}^{+\infty} dx^1\,T^{00}(x^0,x^1)
$
and computing the matrix elements of both terms of this equation on
$\langle \beta_1,a_1|$ and $|\beta_2,a_2\rangle$. Notice however that our
expression differs from that proposed in \refto{Zpoly} which diverges at
infinity much faster than ours.}
$$
<0\mid\Theta(0)\mid a,\beta_1;b,\beta_2>\,=\,-
2\pi\,i \,\delta_{ab}\,M^2 F_{\rm min}(\beta_1-\beta_2)
\quad,
\eqno(FFtrace)
$$
with $F_{\rm min}(\beta)$ given by \(fminpoly). Truncating the spectral
representation of the scaling function $\Phi_1(\rho)$ to this term we have
for the RHS of \(U2)
$$
2\pi\,\int_0^{\infty}\frac{d\beta}{\cosh^4\beta} |F(2\beta)|^2
\,=\,2.2241
\quad,
\eqno(U2approx)
$$
to be compared with the exact result of $U_2$, from \(s),
extracted by the $c$-theorem sum rule (in the limit $n\rightarrow 0$)
$$
U_2\,=\,\frac{20}{9}\,=\,2.2222
\quad.
\eqno(U2ex)
$$
Therefore we see that by keeping only the two-particle approximation of the
correlator, we get an approximation to $U_2$ with a precision of about
one part in $10^3$. However, the careful reader will notice that in this case
the two-particle approximation results in an estimate of this quantity slightly
larger than the exact value. Since the spectral representation of the
scaling function $\Phi_1(\rho)$ involves an infinite series of terms
which are integrals over the modulus squared of its form factors, one may
expect an underestimate of the exact value of $U_2$ from the 2-particle
approximation. Although this is what usually happens in ordinary QFT, the
situation for the ${\rm O}(n)$ (in the limit $n\rightarrow 0$) may be
different.
To see this, let us consider for instance the four-particle FF of the
trace of the energy-momentum tensor
$$
\langle0|\Theta|\beta_1,a;\beta_2,b;\beta_3,c;\beta_4;d\rangle
\quad.
\eqno(FF4)
$$
{}From group theory considerations this has the form
$$
A\,\delta_{ab}\delta_{cd}
+ B\,\delta_{ad}\delta_{bc} +
C\,\delta_{ac}\delta_{bd}
\quad,
\eqno(group)
$$
where the invariant amplitudes $A,B,C$ (functions of
$\beta_1,\ldots,\beta_4$ and $n$) satisfy the functional and
recursive equations of the form factors. For our considerations,
their explicit form is not needed. In order to consider the four-particle
contribution to the scaling function $\Phi_1(\rho)$, we have to take the
modulus squared of this form factor
and sum on the colour indices. The final expression
is given by
$$
(\mid A\mid^2 + \mid B\mid^2 + \mid C\mid ^2) n^2+(A \bar{B} +
B \bar{A} + B \bar{C} + C \bar{B} +C \bar{A} + A \bar{C}) n
\quad,
\eqno(square1)
$$
which may be rewritten as
$$
(\mid A\mid^2 +\mid B\mid^2 +\mid C\mid^2) \,(n^2-n)\, + \,\mid A + B + C
\mid^2 \,n
\quad.
\eqno(square2)
$$
This is clearly positive for $n\geq1$, and in this case we have the usual
positive contribution of this form factor
to the two-point function. But, for $n\to0$,
we need to pick up the linear term in $n$, i.e. the cross term in \(square1),
which has no reason to be positive. As a matter of fact, for $n=1$, which is
the Ising model, we know it is negative, because in that case there is no
4-particle form factor so that $A+B+C=0$. Hence, even though we have not
pursued explicitly the calculation of the above amplitudes $A,B,C$ for the
${\rm O}(n)$ model, one should not be surprised to find that higher particle
contributions will contribute negatively to the correlation functions. This
effect becomes more evident if we consider the zero moment of $\Phi_1(\rho)$
$$
U_0\,=\,\int \Phi_1(\rho)\,d^2\rho
\quad.
\eqno(U0)
$$
This time there are no extra powers in $\rho$ which suppress
the ultraviolet behaviour of the correlation function, and higher form
factors may expected to be important. As shown in the next section, $U_0$
may be computed exactly through the TBA
$$
U_0\,=\,\frac{4\pi^2}{3}\,=\,13.1595
\quad.
\eqno(U0ex)
$$
On the other hand, the two-particle approximation to this quantity
is given by
$$
U_0^{(2)}\,=\,2 \pi\,
\int_0^{\infty} dt \frac{\mid F_{\rm min}(2t)\mid^2}{\cosh^2 t}
\,=\,14.3651
\quad,
\eqno(U0appr)
$$
which is close to the exact result, but slightly larger.

\head{4. TBA and the ratio $C/D$.}
\subhead{4.1 The bulk free energy and the amplitude $U$.}
In this section, we discuss the calculation of the universal amplitude
$U$ in the extensive part of the free energy, defined by
$\ln Z\sim {\cal A}\,n\,UM^2$, where $M=\xi^{-1}$ is the mass of the
${\rm O}(n)$ model, in the limit $n\to0$. This is performed using the
thermodynamic Bethe ansatz (TBA). This method was first applied to
integrable perturbed conformal field theories in a pioneering paper by
Al.~Zamolodchikov\refto{Z1},
who also later\refto{Z2} applied it to the minimal
models perturbed by the $(1,3)$ operator. Since our calculation is a
rather simple generalization of his, we shall only sketch the argument,
and refer the reader to Zamolodchikov's papers for the rather heavy
details. Thus, in this section, we make no attempt to be self-contained.

In the TBA approach, one considers the massive theory as a quantum field
theory in $1+1$ dimensions, at finite temperature $R^{-1}$. Denoting the
reduced free energy per unit length of this system by $E(R)$,
dimensional analysis implies that this has the form $E(R)=(2\pi/R)F(r)$,
where $r=MR$. As $r\to0$, the scaling function has the form $F(0)+ O(r^2) +
{\rm singular\ terms}$. The terms proportional to $F(0)$ dominates in
the high-temperature limit $R\to0$ and gives the central charge of the
conformal field theory. The $O(r^2)$ piece gives a term extensive in
both the temperature and the spatial size of the system, and is
therefore the extensive part of the free energy when we view the system
as a classical system in two dimensions, which is what we wish to
compute.

The TBA proceeds by enumerating the allowed states in a large spatial
box, labelled by the rapidities of the particles, consistent with the
fact that the wave functions must change by a factor of the $S$-matrix
whenever two particles are exchanged. The fraction of these states which
are actually occupied is then determined by minimizing the free energy.
For a simple theory with only one type of excitation of mass $M$,
the result is that
$$
E(R)=-M\int L(\beta)\cosh\beta{d\beta\over2\pi}
\quad,\eqno(4a)
$$
where
$$
L(\beta)=\ln(1+e^{-\epsilon(\beta)})
\quad,\eqno(4b)
$$
is the solution of the integral equation
$$
-MR\cosh\beta + \epsilon(\beta) +(\phi*L)(\beta)=0
\quad,\eqno(4c)
$$
where $*$ denotes a convolution, and $\phi(\beta)=-i(d/d\beta)
\ln S(\beta)$.

Zamolodchikov showed\refto{Z1} that, in the limit $R\to0$, it is
sufficient to consider the `kink' solution of the simpler integral
equation
$$
-e^\beta+\epsilon^{\rm kink}(\beta)+(\phi*L^{\rm kink})(\beta)=0
\quad,\eqno(4d)
$$
and that
$$
F(r)\sim -{1\over 2\pi^2}\int L^{\rm kink}(\beta)
e^\beta d\beta
    -{r^2\over8\pi^2}\int{dL^{\rm kink}(\beta)\over
d\beta}e^{-\beta}d\beta+ {\rm singular\ terms}
\quad.\eqno(4e)
$$
Furthermore, the integral in the second term may be evaluated easily
by examining the behaviour of \(4d) as $\beta\to-\infty$, and requiring
that the $O(e^\beta)$ terms cancel between the first and third terms.
Thus if $\phi(\beta)\sim Ae^\beta$ as $\beta\to-\infty$, then the value
of the integral in the second term of \(4e) is $2\pi/A$.

In \ref{Z2}, Zamolodchikov considered the case of the unitary minimal
models, labelled by an integer $k$, which correspond to the critical
continuum limit of the lattice RSOS models, perturbed by the operator
$(1,3)$. Although the ${\rm O}(n)$ model and the minimal model with
$n=2\cos(\pi/(k+1))$ have the same value of the central charge, their
operator content is certainly different. However, within the sector of
operators of the type $(1,2s+1)$, generated by repeated operator product
expansions of $(1,3)$ with itself, they are the same. Thus, to any order
in perturbation theory, the correlation functions of the two theories
should agree, and therefore so should such quantities as the mass gap
and ground state energy. We assume that this will also hold
non-perturbatively. In principle, these theories could differ at the
non-perturbative level, but we see no physical reason for this to
happen. Zamolodchikov argued that, although there is only one physical
massive excitation in the RSOS model, in order to count correctly the
states it is necessary to introduce pseudo-excitations which carry no
energy but do enter into the TBA equations. For the $k$th minimal model
these excitations correspond to the vertices of the $A_{k-1}$ Dynkin diagram,
labelled by $a=1,\ldots,k-1$. The
physical particle corresponds to $a=1$. The kink integral equation \(4d)
becomes
$$
-\delta_{a1}+\epsilon_a^{\rm kink}(\beta)+\sum_bl_{ab}(\phi*L_b^{\rm
kink})(\beta)=0
\quad,\eqno(4f)
$$
where $l_{ab}$ is the incidence matrix of the diagram. Thus, by looking
at the $\beta\to-\infty$ and using $A=2$ we find the following system
of equations for the integrals $I_a=\int(dL_a^{\rm
kink}/d\beta)e^{-\beta}d\beta$:
$$
\eqalign{
I_2&=\pi\cr
I_1+I_3&=0\cr
I_2+I_4&=0\cr
\vdots\cr
I_{k-2}&=0\quad.\cr}
\eqno(4g)
$$
Strictly speaking, these equations make sense only when $k$ is an
integer. However, we may formally find the solution for general $k$,
which is
$$
I_j=\pi{\sin\big((j-k)\pi/2\big)\over\sin(k\pi/2)}
\quad.\eqno(4h)
$$
Since only $a=1$ has mass, the term we want is proportional to
$I_1=\pi\cot(k\pi/2)$. Putting all the factors together, we find that
$$
F(r)=F(0)-{r^2\over8\pi}\cot(k\pi/2)+{\rm power\ series\ in\ }r^{4/(k+2)}
\quad.\eqno(4i)
$$
Note that when $k$ is an odd integer, the extensive part of the free
energy vanishes, as found by Zamolodchikov: this is interpreted as being
a consequence of (fractional) supersymmetry. For $k=2p$ the coefficient
diverges, but, since the overall result is finite, there must be a
cancellation of this leading term against one of the singular terms,
resulting in logarithmic behaviour. This comes from
$$
\lim_{k\to2p} -{1\over8\pi}{2/\pi\over k-2p}\left(r^2-
r^{4(p+1)/(k+2)}\right)=-{2\over k+2}\left({r\over2\pi}\right)^2\ln r
\quad,\eqno(4j)
$$
in agreement with (2.22) of \ref{Z2}.

However, we are interested in the limit $n\to0$, corresponding to
$k\to1$. In that limit we then find for the free energy per unit area
${\cal A}^{-1}\ln Z\sim (\pi/8)(k-1)M^2$. Since $dk/dn|_{n=0}=2/\pi$,
we then have the final result that
$$
U=\frac14
\quad.\eqno(4k)
$$
\vfill\eject
\subhead{4.2 The ratio $C/D$.}
We are now in a position to calculate the amplitude $D$ defined in
\sec{2}. From \(s,u,4k) we find that
$U_2/U_0=5/3\pi^2$ and so, from \(v),
$$
D={5\over 6\pi^{\frac32}}\xi_0^2
\quad.\eqno(4l)
$$

Now consider the amplitude $C$, which, by \(ab), is proportional to the
ratio of the second and zeroth moments of the scaling function $\Phi_2$.
At the same level of approximation, we should neglect all $m$-particle
intermediate states with $m>2$. But, since the spin operator couples
only to odd $m$, this implies that we need include only the single
particle intermediate state, which is equivalent to assuming that
$$
\Phi_2(\rho)=\int{e^{iq\cdot\rho}\over q^2+1}{d^2q\over(2\pi)^2}
\quad.\eqno(4m)
$$
Note that this is {\it not} the same as assuming the Ornstein-Zernicke
(free-particle) form for the correlation function, which would imply that
$\eta=0$ and $\gamma=2\nu$. In writing \(4m) for the scaling
function, we have already extracted in \(ab) the dependence coming from
the exact value $\gamma$. From \(4m) we find the moment ratio
$V_2/V_0\approx4$, so that finally, from \(ac), we obtain
one of our main results
$$
C/D\approx{24\pi^{\frac32}\over5}{\Gamma(\ffrac{43}{32})\over
\Gamma(\ffrac{91}{32})}\approx 13.70
\quad.\eqno(4n)
$$
It is difficult to estimate the errors in this calculation, which arise
from the neglect of the 3- and higher particle intermediate states in
the spin-spin correlation function. We would expect the error to be
larger for the zeroth moment $V_0$ than for $V_2$. On the grounds of
phase space alone, one would expect the $m$-particle state to give a
relative contribution of the order of $\int e^{-mMr}d^2r\sim m^{-2}$
to the zeroth moment. This will be modified by a factor of
$m^{-m(m-1)/4}$ due to the softening of the multi-particle branch points
discussed in \sec{3}, leading to an expected error of a few percent in
$V_0$. This estimate, however, does not take into account
the further
suppression of the small $r$ region due to the small value of the
magnetic scaling index $\eta$.
\vfill\eject
\subhead{4.3 Connection with the conformal limit.}
There is more information to be gained from the TBA. So far the energy
operator ${\cal E}(r)$ has appeared only in the combination $(2\pi/\nu)
(x_c-x)\,{\cal E}(r)$, as the trace $\Theta(r)$ of the stress tensor. But if we
fix on a definite normalization of ${\cal E}(r)$, for example, that given
naturally by its conformal limit, then there is a definite relation
between the temperature-like variable $(x_c-x)$ and the mass $M$.
As shown by Al.~Zamolodchikov\refto{Z1,Z2}, this may be determined by
comparing the corrections to the TBA with the results of a perturbative
analysis. This analysis has been carried through already\refto{FSZ} for the
${\rm O}(n)$ model
for a slightly different situation than that considered above, namely,
when the non-contractible loops which wind around the imaginary time
direction carry a factor 2 rather than $n$. However, this change of
boundary condition should not affect the relationship we wish to
determine.

With this boundary condition on the cylinder, the effective UV central
charge is $\tilde c=1$, rather than vanishing in the limit $n\to0$. This
may be traced to the existence of a negative dimension operator with
scaling dimensions $(\Delta_0,\overline\Delta_0)$, such that
$$
\tilde c\,=\,c-12(\Delta_0+\overline\Delta_0)
\eqno(effective)
$$
so we see that $\Delta_0=\overline\Delta_0=-\frac1{24}$, when $n=0$.
Noticing that these are the scaling dimensions of the operator
$\varphi_{2,3}$ in the Kac table, we may interpret this choice of boundary
conditions as arising from the insertion of this operator at each end of
the cylinder. Thus the perturbative expansion of the ground state
energy begins as\refto{Cardy}
$$
E(R)=-{\pi\over6R}+R(x_c-x)\left(2\pi\over
R\right)^{2/3}C^{\cal E}_{(2,3),(2,3)}+\cdots
\quad,
\eqno(perturbative)
$$
where $C^{\cal E}_{(2,3),(2,3)}$ is the structure constant of the three point
function $\langle {\cal E}\varphi_{2,3}\varphi_{2,3}\rangle$.

In our problem it is natural to normalize the energy operator so that
as $r\to0$,
$$
\langle {\cal E}(r) {\cal E}(0)\rangle\sim n/r^{4/3}
\quad,
\eqno(Enorm)
$$
so that, with periodic boundary conditions, the free energy will be
order $n$ as required. This means that we should take ${\cal E}=\sqrt
n\varphi_{1,3}$, where $\varphi_{1,3}$ is normalized in the conventional
fashion.
As a result the structure constant is
$$
C_{(2,3),(2,3)}^{\cal E}=\lim_{k\rightarrow 1} \sqrt{n(k)}
C_{(2,3),(2,3)}^{(1,3)}\,=\,
\frac{3^{9/4}}{16\pi^3}\frac{\Gamma^8(2/3)}{\Gamma(4/3)}\,
\,=\,0.302248
\quad.
\eqno(stucture)
$$
Note that this has a finite limit as $n\to0$.

Defining the scaling function
$\tilde c(MR)=-(6R/\pi)E(R)$ we therefore have
$$
\tilde c(MR)\,=\,1\,-\,12(2\pi)^{-1/3}\,(0.302248)\,(x_c-x)\,R^{4/3}+
\cdots
\quad.
\eqno(per)
$$
On the other hand, in \ref{FSZ} $\tilde c(MR)$ has been
determined as the numerical solution of the TBA equations, and the first
terms of its expansion are given by
$$
\tilde c(MR)\,=\,1\,-\,0.4454536\, (MR)^{4/3}+\cdots
\quad.
\eqno(TBAcentral)
$$
Comparing these, we obtain the number which links the physical mass $M$ to
the scale of temperature, given our choice of normalization of ${\cal E}$
$$
(x_c-x)\,=\,\kappa \,M^{4/3}
\quad,
\eqno(universal)
$$
where
$\kappa\approx0.226630$.
Moreover, using our parametrization
$M=\xi_0^{-1}(1-x/x_c)^{3/4}$,
we also see that
$$
x_c\,=\,\kappa\,\xi_0^{-4/3}
\quad.
\eqno(xcxi)
$$
This relation will be used in section 5 to fix the asymptotic behaviour of
the structure factor.

\head{5. The Structure Factor of Self-Avoiding Loops}

An important measure of self-avoiding walks, which is, in principle,
accessible by
light scattering experiments, is the structure factor
of $N$-step loops
$$
{\cal S}_N(q)=N^{-1}\left|\sum_{i}\,\langle e^{iq\cdot r_i}
\rangle\right|^2=N^{-1}\sum_{i,j}\langle e^{iq\cdot (r_i-r_j)}
\rangle
\quad.
\eqno(structdefin)
$$
${\cal S}_N(q)$ is a positive definite quantity that plays the role of
generating function for all moments of the energy-energy correlator.
In the continuum limit we have
$$
n\sum_N N p_N{\cal S}_N(q)x^{N-2} = a_0
\int d^2x e^{iq\cdot x}\langle {\cal E}(x) {\cal E}(0)\rangle
\quad,
\eqno(gen1)
$$
and ${\cal S}_N(q)$ may be expressed as
$$
{\cal S}_N(q)\,\,\mathop\simeq_{N\rightarrow\infty}\,\,
\,N F(y)\,=\,N \left(1-\frac{y^2}{2}+
\ldots\right)
\quad,\eqno(scaling)
$$
where $y$ is the scaling variable given in terms of the mean square radius of
gyration of loops
$$
y^2\,=\,q^2\,\langle R^2_g\rangle_N
\quad.
\eqno()
$$
As we are going to show, the scaling function $F(y)$ may be analyzed in great
detail by means of two different methods, the form factor approach and
the conformal field theory. By the former we are able to arrive at quite
an accurate estimate of the scaling function $F(y)$ (at least for moderate
values of the scaling variable $y$), whereas, by the latter we may
extract its asymptotic behaviour.
\subhead{5.1 Structure Factor in the Two-Particle Approximation}
A precise determination of $F(y)$ may be obtained in terms of the
two-particle form factor as follows. Expand initially both terms
in \(gen1) in power series of $q$
$$
n\sum_{N,p}N p_N {\cal S}_{N,2p} \,q^{2p} \,x^{N-2} = a_0 \,
\sum_{p=0}^{\infty} \frac{1}{(2p)!} \,
\int d^2x (iq\cdot x)^{2p}\,\langle {\cal E}(x) {\cal E}(0)\rangle
\quad.
\eqno(gen2)
$$
(the odd terms in the series vanish by parity). The key point is that we
may compute {\it exactly} the first two coefficients of the series by
exploiting
the TBA and the $c$-theorem sum rule. As the other terms are related to the
higher moments of the energy-energy correlation function
$$
\int d^2x \,|x|^{2p}\,\langle {\cal E}(x) {\cal E}(0)\rangle
\quad,
\eqno(moments)
$$
they will be approximated with quite high accuracy by retaining only the
two-particle contribution. The reason is that the higher powers $|x|^{2p}$
emphasize more the large distance scales of the correlation function,
and on the contrary suppress its short distance singularity. Hence they
improve substantially the ultraviolet convergence of the integral and
the spectral representation of the correlator $\langle {\cal E}(x){\cal
E}(0)\rangle$ is
effectively truncated to the two-particle term\footnote{$^1$}{As a matter
of fact, we have already seen in sect. 3 that the two-particle contribution
provides an accurate estimate for the second moment. Hence, we expect
the precision obtained by the two-particle contribution to increase for the
higher moments.}.

Let us then write the right hand side of \(gen2) as
$$
a_0\, \left(I_1 -\frac{1}{2} I_2 + I_3\right)
\quad,\eqno(split)
$$
where
$$
\eqalign{
& I_1\,=\, \int d^2x\, \langle {\cal E}(x) {\cal E}(0)\rangle \cr
& I_2\,=\, \int d^2x (q\cdot x)^2 \, \langle {\cal E}(x) {\cal E}(0)\rangle \cr
& I_3 \,=\, \sum_{p=2}^{\infty} (-1)^p\,\int d^2x \frac{1}{(2p)!}
(q\cdot x)^{2p} \,\langle {\cal E}(x) {\cal E}(0)\rangle \quad.\cr}
\eqno(Identification)
$$
As discussed in \sec{2}, $I_1$ and $I_2$ are exactly obtained in terms of the
extensive part of the free energy and by the $c$-theorem:
$$
\eqalign{
& I_1 \,=\, n\frac{2\nu(2\nu-1)}{(x_c-x)^2} M^2 \,U \cr
& I_2\,=\, n\frac{q^2}{2} \frac{\nu^2}{3\pi (x_c-x)^2} \,c'(0)
\quad.\cr}
\eqno(I1-2)
$$
$I_3$ is related to the higher moments of
$\langle{\cal E}(x){\cal E}(0)\rangle$, and,
computed in the two-particle approximation, it is given by
$$
I_3\,=\,\frac{nM^2 \nu^2}{2\pi (x_c-x)^2}
\sum_{p=2}^{\infty} (-1)^p \frac{q^{2p}}{(4 M^2)^p} \,{\cal I}_{2p}
\quad,
\eqno(I3)
$$
where
$$
{\cal I}_{2p}\,=\,\int_0^{\infty} dt \,\frac{\mid F_{\rm min}(2t)\mid^2}
{\,\,\,(\cosh t)^{2+2p}}
\quad.
\eqno(I2p)
$$
The numerical evaluation of the first ${\cal I}_{2p}$ is reported in Table I.
They vanish asymptotically as
$$
{\cal I}_{2p}\,\,\mathop\simeq_{p\rightarrow\infty}\,\,
\frac{{\sqrt\pi}}{4}\,\frac{\Gamma\left(p-\frac{3}{4}\right)}
{\Gamma\left(p+\frac{3}{4}\right)}\,\,\Xi^2(0)
\quad,
\eqno(asyI2p)
$$
where $\Xi(0)$ is a constant defined in the appendix.

To obtain ${\cal S}_N(q)$ from the RHS of \(gen2) we need to express $M$ as
$\xi_0^{-1}(1-x/x_c)^{3/4}$ and to take the inverse Laplace transform. The
result is
$$
\eqalign{
&
\frac{{\cal S}_{N,2}}{{\cal S}_{N,0}} \,=\, - \frac{5}{12 \pi {\sqrt \pi}}
\,\frac{\Gamma(N)}{\Gamma\left(N-\frac{3}{2}\right)} \,\xi_0^2 \cr
&\frac{{\cal S}_{N,2p}}{{\cal S}_{N,0}} \,=\, (-)^p \frac{3}{2 {\sqrt \pi}}
\,\frac{{\cal I}_{2p}}{2^{2p}}
\,\frac{\Gamma\left(N+\frac{3(p-1)}{2}\right)}
{\Gamma\left(N-\frac{3}{2}\right)\,
\Gamma\left(\frac{3p+1)}{2}\right)} \,\xi_0^{2p}\quad. \cr}
\eqno(ratiostr)
$$
The non-universal factor $\xi_0$ as in \(ratiostr) may be eliminated by using
both the scaling form of ${\cal S}_N(q)$ (\(scaling)) and the corresponding
definition of $\langle R_g^2\rangle_N$ obtained by
${\cal S}_{N,2}/{\cal S}_{N,0}$
$$
\langle R^2_g\rangle_N\,\,\mathop\simeq_{N\rightarrow \infty}\,\,
\frac{5}{6\pi\sqrt{\pi}}\,N^{2\nu}\,\xi_0^2
\quad,
\eqno(R2)
$$
in agreement with \(4l). With this substitution, the scaling function $F(y)$
is given by
$$
F(y)\,=\,\left(1-\frac{y^2}{2}+\sum_{p=2}^{\infty} B_{2p}\, y^{2p}
\right)
\quad,
\eqno(final)
$$
where
$$
B_{2p}\,=\, (-1)^p \frac{3}{2 {\sqrt \pi}}\,
\frac{{\cal I}_{2p}}{\Gamma\left(\frac{3 p+1}{2}\right)}
\,\left(\frac{3\pi {\sqrt\pi}}{10}\right)^p
\quad.
\eqno(coefficients)
$$
Their values for $p\leq20$ are reported in Table I.
The resulting series is convergent for all complex $y$, and in
particular yields a highly precise determination of the scaling function $F(y)$
for small values of $y$. Although the higher order coefficients become
more and more precisely determined by the 2-particle approximation, this
does not mean that this approximation captures the correct large $y$
behaviour of the scaling function. Indeed, it is straightforward to see
that, within the 2-particle approximation, this is determined by the
large $\beta$ behaviour of $F_{\rm min}(\beta)$, which
leads to a power law $y^{-1/2}$. A more accurate way of finding the
large $y$ behaviour is through conformal field theory.
\subhead{5.2 Asymptotic Behaviour of the Structure Factor from CFT}
In order to analyze the asymptotic behaviour of $F(y)$ in the limit
$y\rightarrow\infty$, let us first write the RHS of \(gen2) as
$$
A(q)\,=\, \frac{2\pi a_0}{q^2}\,
\int_0^{\infty} dt\,t\,J_0(t)\,\langle {\cal E}(t/q)\,{\cal E}(0)\rangle
\quad.
\eqno(bessel)
$$
($J_0(t)$ is the Bessel function). In the limit $q\rightarrow \infty$
only the short distance scales of the correlation function are important and
therefore $\langle {\cal E}(t/q)\,{\cal E}(0)\rangle$ is determined by the
operator product expansion:
$$
\langle {\cal E}(r)\,{\cal E}(0)\rangle
\,\,\mathop\simeq_{r\rightarrow 0}\,\,
r^{-4/3}\,+\,{\cal C}\, \langle {\cal E}(0)\rangle\,r^{-2/3}\,+\,\ldots
\quad.
\eqno(OPE)
$$
$\langle {\cal E}(0)\rangle$ is the vacuum expectation value of the energy
operator determined by TBA
$$
\langle {\cal E}(0)\rangle\,=\,-\,\frac{\nu}{2\pi\,(x_c-x)}\,
\langle \Theta(0)\rangle\,=\,-\,\frac{3}{32\pi}\frac{M^2}{(x_c-x)}
\quad,
\eqno(vevE)
$$
whereas ${\cal C}$ is the structure constant of the CFT algebra. To compute it,
we have to use the formulas of \ref{DF2} and take into account
the fact that we have chosen to normalize the 2-point function of the
energy operator ${\cal E}(r)$ to $n$ rather than unity. Therefore ${\cal C}$
is given by the limit
$$
{\cal C}\,=\, \lim_{k\to 1} \sqrt{n(k)} \,\,
{\cal C}_{(1,3),(1,3)}^{(1,3)}(k)\,=\,
6\,\sqrt {6\pi}\,
\left(
\frac{\Gamma\left(\frac{2}{3}\right)}{\Gamma\left(\frac{1}{3}\right)}
\right)^{\frac{9}{2}}
\quad.
\eqno(strucconst)
$$
Inserting \(OPE) into \(bessel), the large $q$ behaviour of $A(q)$
is given by
$$
\eqalign{
A(q) \,&\mathop\simeq_{q\rightarrow\infty}\, A^{\left(2/3\right)}
\,+\,A^{\left(4/3\right)}\,+\,\ldots \,\cr
&=\,\frac{2\pi\,a_0\,J_1}{q^{2/3}} \,-\,
\frac{3\,a_0\,\,M^2\,{\cal C}\,J_2}{32\pi\,(x_c-x)}\,\frac{1}
{q^{4/3}} \,+\,\ldots
\quad,\cr}
\eqno(1appr)
$$
where
$$
\eqalign{
& J_1\,=\,\int_0^{\infty} dt\, t^{-1/3}\,J_0(t)\,=\,2^{1/3}\,
\frac{\Gamma\left(\frac{2}{3}\right)}{\Gamma\left(\frac{1}{3}\right)}\cr
& J_2\,=\,\int_0^{\infty} dt\, t^{1/3}\,J_0(t)\,=\,2^{-1/3}\,
\frac{\Gamma\left(\frac{1}{3}\right)}{\Gamma\left(\frac{2}{3}\right)}
\quad\cr}
\eqno(b1/3)
$$
To obtain the scaling function $F(y)$ we still have to perform an inverse
Laplace transform on $A(q)$. However the leading order $A^{(2/3)}$ of $A(q)$
does not contain any dependence on $(x_c-x)$ and therefore in the asymptotic
expansion of $F(y)$,
$$
F(y)\,\mathop\simeq_{y\rightarrow\infty}\, \frac{a}{y^{2/3}}\,+\,
\frac{b}{y^{4/3}}\,+\,\ldots
\quad,
\eqno(fasy)
$$
the coefficient $a$ vanishes identically! Hence the scaling function
$F(y)$ decreases at infinity faster than expected by a power-counting
argument. As we will discuss at the end of this section, the validity of
this result is not restricted to two-dimensional self-avoiding loops but, on
the contrary, is quite general. In two dimensions, however, we may use the
exact value of the structure constant ${\cal C}$ to extract the universal
coefficient $b$ in \(fasy). In fact, expressing $M$ as
$\xi_0^{-1}(1-x/x_c)^{3/4}$ and making an inverse Laplace transform
on $A^{(4/3)}$ we have
$$
Np_N{\cal S}_N(q)\mathop\sim_{q,N\to\infty}\,
-\,{3\sigma a_0\over32\pi}{\cal C}J_2
\xi_0^{-2}x_c{N^{-3/2}\mu^N\over\Gamma(-\ffrac12)}\,\,q^{-4/3}
\quad.
\eqno(inLap1)
$$
Normalizing this quantity to the corresponding expression for $q\to 0$, i.e.
$$
Np_N{\cal S}_N(0)\mathop\sim_{N\to\infty}\,
{3\sigma a_0\over 16}\xi_0^{-2}{N^{-1/2}\mu^N\over\Gamma(\ffrac12)}
\quad,
\eqno(norLap1)
$$
(in agreement with \(e,4k),)
and using the relation $\langle R^2\rangle_N\sim5\xi_0^2N^{3/2}/6\pi^{3/2}$
from \(4l), we see that the coefficient $b$ in \(fasy) depends on the
combination $\kappa=x_c\xi_0^{4/3}$, which was computed in \sec{4.3}.
The appearance of this quantity in this context is not surprising, since
we are considering the ultraviolet behaviour of ${\cal S}(q)$, which is
normalized at $q=0$.
Putting all this together, we obtain the final result
$$
F(y)\,\mathop\simeq_{y\rightarrow\infty}\, \frac{b}{y^{4/3}}\,+\,\ldots
\quad,
\eqno(fasy2)
$$
where $b$ is the universal constant
$$
\eqalign{
&b\,=\,\frac{\kappa}{4\pi}\,J_2 \,
\left(\frac{5}{6\pi{\sqrt\pi}}\right)^{\frac{2}{3}}
\,{\cal C} \,=\cr
&\,\,=\,\kappa\,{3^{\frac56}\,5^{\frac23}\over(2\pi)^{\frac32}}\,
\left(\frac{\Gamma\left(\frac{2}{3}\right)}{\Gamma\left(\frac{1}{3}\right)}
\right)^{\frac{7}{2}}\,=\, 9.65065\,\times 10^{-3}\quad.\cr}
\eqno(buniversal)
$$

Returning to the question of the leading coefficient $a$ of $F(y)$ which
vanishes, it is easy to show that this result is quite general and may be
simply justified by a physical argument. In fact, in $d$ dimensions
the leading behaviour of the generating function of
${\cal S}_N(q)$ decreases as $q^{d-2/\nu}$, but with a corresponding
coefficient which is independent of $(x_c-x)$. The sub-leading term goes
in any dimension as $q^{-1/\nu}$ but the coefficient in front depends
on the contrary on $(x_c-x)$. Therefore, the scaling function $F(y)$
always decreases at infinity as $y^{-1/\nu}$. This is what one expects for an
object whose fractal dimension is $1/\nu$. It is easy to see that the same
power law also rules the large $q$ behaviour of the structure factor of a
linear polymer whose generating function is given by
$$
\int e^{iq(r-r')}\,\langle s(0) {\cal E}(r) {\cal E}(r')s(r_1)\rangle_c
\,d^dr\,d^dr'\,d^dr_1
\quad.
\eqno(linear)
$$
Since this is a connected correlation function, in the limit
$q\rightarrow\infty$ the leading term comes from the ${\cal E}$ term in the OPE
of ${\cal E}(r) {\cal E}(r')$ rather than from the unit operator. Hence the
decrease of $F(y)$ as $y^{-1/\nu}$ is consistent with the idea that linear
and loop polymers have the same fractal dimension on small scales.

\head{6. Area of loops and the current-current correlation function.}

In this section we show how the amplitude governing the mean square area
of
self-avoiding closed loops is related to the second moment of a
current-current correlation function in the ${\rm O}(n)$ model. A
summary of this calculation was given in \ref{CG}. Consider a loop of
$N$ steps. It has been argued that the mean area $\langle a\rangle_N$
of such loops should
behave asymptotically as $EN^{2\nu}$, and, more generally, that the
moments $\langle a^p\rangle_N$ of the area should behave as
$E^{(p)}N^{2p\nu}$, where the amplitude combinations $E^{(p)}/D^p$ are
expected to be universal. In this paper we focus on the case $p=2$.

Any given loop may be assigned two possible orientations. Consider a
given orientation, and let $J_\mu^{\rm lat}(r)$ be a vector of unit
magnitude on the link $r$ in the direction of the orientation. Then the
{\it signed} area of a given loop is
$$
a=\ffrac12\sum_r\epsilon_{\mu\nu}r_\mu J_\nu^{\rm lat}(r)
\quad.\eqno(6a)
$$
Of course, this quantity averages to zero when summed over orientations,
but its mean square is non-zero. For a given loop
$$
a^2=\ffrac14\sum_{r,r'}\Big(r_\mu r'_\mu J_\nu^{\rm lat}(r)J_\nu^{\rm
lat}(r')-
r_\mu r'_\nu J_\nu^{\rm lat}(r)J_\mu^{\rm lat}(r')\Big)
\quad.\eqno(6b)
$$
Using the results $\sum_rJ_\nu^{\rm lat}(r)=0$ and
$\sum_r r_\nu J_\nu^{\rm lat}(r)=0$, which follow from the fact that
$J_\nu$ is conserved, \(6b) may be rewritten as
$$
a^2=\ffrac14\sum_{r,r'}\Big(-\ffrac12(r-r')^2J_\nu^{\rm lat}(r)
J_\nu^{\rm lat}(r')+(r_\mu-r'_\mu)(r_\nu-r'_\nu)J_\mu^{\rm lat}(r)
J_\nu^{\rm lat}(r')\Big)
\quad.\eqno(6c)
$$
This is for one loop. If we now average over all possible oriented
loops,
weighted by a factor $x^N$ for each loop, we find that
$$\eqalign{
2n{\cal N}_s\sum_Np_N&\langle a^2\rangle_Nx^N\cr
&=\ffrac14\sum_{r,r'}\Big(-\ffrac12(r-r')^2\langle J_\nu^{\rm lat}(r)
J_\nu^{\rm lat}(r')\rangle+(r_\mu-r'_\mu)(r_\nu-r'_\nu)\langle
J_\mu^{\rm lat}(r)J_\nu^{\rm lat}(r')\rangle\Big)\quad,\cr}
\eqno(6d)
$$
where the correlation functions on the right hand side are evaluated in
a {\it complex} ${\rm O}(n)$ lattice model, with hamiltonian
$H_2=-x\sum_{\rm nn}s_a^*(r)s_a(r')$, in the limit $n\to0$. The current
is the one which generates the ${\rm U}(1)$ transformations $s_a\to
e^{i\alpha}s_a$. An explicit expression for $J_\mu^{\rm lat}(r)$ was
given in this model in \ref{CG}.

\(6d) is now ready for taking the continuum limit. (Doing this earlier on
\(6b) leads to erroneous results.) To do this, we simply let
$\sum_rJ_\mu^{\rm lat}(r)\to\int J_\mu(r)d^2r$, so that the right hand
side of \(6d) becomes
$$
\ffrac14{\cal A}\int(-\ffrac12r^2\delta_{\mu\nu}+r_\mu r_\nu)
\langle J_\mu(r)J_\nu(0)\rangle d^2r
\quad.\eqno(6e)
$$
Since $J_\mu$ is conserved, by rotational symmetry the correlation
function on the right hand side has the form
$$
\langle J_\mu(r)J_\nu(0)\rangle=(\partial_\mu\partial_\nu-\delta_{\mu\nu}
\partial^2)G(r)
\quad,\eqno(6f)
$$
where $G$ is a scalar. Substituting this into \(6e) and integrating by
parts, the right hand side is simply ${\cal A}\int Gd^2r$. On the other
hand, using the same method, one may show that $\int r^2\langle J_\mu(r)
J_\mu(0)\rangle d^2r=-4\int Gd^2r$. Thus
$$
2n\sum_Np_N\langle a^2\rangle_Nx^N
=-\ffrac14a_0\int r^2\langle J_\mu(r)J_\mu(0)\rangle d^2r
\quad.\eqno(6g)
$$
Now, since the current has unit scaling dimension, its two-point
function has the scaling form
$$
\langle J_\mu(r)J_\mu(0)\rangle=-n\xi^{-2}\Phi_J(r/\xi)
\quad.\eqno(6h)
$$
Moreover, the normalization of this current is fixed by the requirement
that the ends of a self-avoiding walk, which in the complex ${\rm O}(n)$
model are represented by insertions of $s_1^*$ and $s_1$, are
respectively unit sources and sinks for $J_\mu$. Therefore the scaling
function $\Phi_J$, like that of the correlation function of the trace of
the stress tensor, is completely universal with no metric factors. It
follows that the right hand side of \(6g) has the form
$\frac14na_0U_J\xi_0^2(1-x/x_c)^{-2\nu}$, where $U_J=\int\rho^2\Phi_J
(\rho)d^2\!\rho$. Thus
$$
p_N\langle a^2\rangle_N\sim\ffrac18\sigma a_0\,\xi_0^2\,U_J\,{N^{2\nu-1}
\over\Gamma(2\nu)}\,\mu^N
\quad.\eqno(6ha)
$$
Using $p_N\sim BN^{-2\nu-1}\mu^N$ and the relation \(r) for $BD$, we
then find that
$$
\langle a^2\rangle_N\sim{4\pi^2\over5}\,{D\,U_J\xi_0^2\over\Gamma(2\nu)}
\,N^{4\nu}
\quad.\eqno(6hb)
$$
Finally, using the exact relation \(4l) for $D$, we obtain
$$
{\langle a^2\rangle_N\over \langle R^2\rangle_N^2}
={E^{(2)}\over D^2}={48\pi^3\over 25}\,U_J
\quad.\eqno(6hc)
$$

Now let us see how we might compute the current-current correlation
function, and hence $U_J$,
through the form factor approach. Since the current is an
${\rm O}(n)$ singlet, it couples only to the even particle sector. The
simplest non-trivial form factor is therefore
$$
\langle 0|J_\mu(x^1)|\beta_1,a_1,+;\beta_2,a_2,-\rangle
=e^{iM(\sinh\beta_1+\sinh\beta_1)x^1}
\langle 0|J_\mu(0)|\beta_1,a_1,+;\beta_2,a_2,-\rangle
\quad.\eqno(6i)
$$
In the above $\pm$ denotes the ${\rm U}(1)$ charge of the particle, and
$a_j$ its ${\rm O}(n)$ colour index. It is simpler to work with the
light-cone combinations $J^{\pm}=J_0\pm J_1$. Then, by ${\rm O}(n)$
symmetry, Lorentz covariance and current conservation,
$$
\langle 0|J^\pm(0)|\beta_1,a_1,+;\beta_2,a_2,-\rangle
=\delta_{a_1a_2}(e^{\pm\beta_1}-e^{\pm\beta_2})F(\beta_{12})
\quad.\eqno(6j)
$$
It is straightforward to check that
this satisfies the conservation condition $q^+J^-+q^-J^+=0$, with
$q^\pm=M(e^{\pm\beta_1}+e^{\pm\beta_2})$. In fact the Watson equations
for this form factor then show that the simplest solution is to take
$F(\beta)=dF_{\rm min}(\beta)$, where $d$ is a normalization constant,
and $F_{\rm min}$ was defined in \sec{3}. The normalization is fixed by
crossing \(6j) to find
$$
\langle\beta_1,a_1,-|J^\pm(x^1)|\beta_2,a_2,-\rangle=
-d\,\delta_{a_1a_2}(e^{\pm\beta_1}+e^{\pm\beta_2})F_{\rm min}(i\pi-\beta_{12})
e^{-iM(\sinh\beta_1-\sinh\beta_2)x^1}
\quad.\eqno(6k)
$$
Thus the matrix element of the total ${\rm U}(1)$ charge is
$$\eqalign{
\langle\beta_1,a_1,-|\int &J^0(x^1)dx^1\,|\beta_2,a_2,-\rangle\cr
&=-2\pi\,d\,\delta_{a_1a_2}(\cosh\beta_1+\cosh\beta_2)F_{\rm
min}(i\pi-\beta_{12})\delta(M(\sinh\beta_1-\sinh\beta_2))\quad.\cr}
\eqno(6l)
$$
The left hand side is simply $-2\pi\delta_{a_1a_2}\delta(\beta_1-\beta_2)$.
Therefore, using the fact that\break $F_{\rm min}(i\pi)=1$, we find
$$
d=\ffrac12M
\quad.\eqno(6m)
$$
In the 2-particle approximation, then, the current-current correlation
function is
$$
\langle J_\mu(r)J_\mu(0)\rangle_2=
{nM^2\over4}\int{d\beta_1\over2\pi}{d\beta_2\over2\pi}
(e^{\beta_1}-e^{\beta_2})(e^{-\beta_1}-e^{-\beta_2})
|F_{\rm min}(\beta_{12})|^2e^{-Mr(\cosh\beta_1+\cosh\beta_2)}
\quad.\eqno(6n)
$$
The factor of $n$ comes from the sum over colour indices in the
intermediate state. Note that there is no factor of $1/2!$
since the particles have opposite ${\rm U}(1)$ charge and so are not
identical.

This leads to the estimate for $U_J$ in this approximation
$$
U_{J,2}=-{1\over2\pi}\int_0^\infty{(1-\cosh2\beta)|F_{\rm min}(2\beta)|^2
\over\cosh^4\beta}d\beta
\quad.\eqno(6o)
$$
The integral may be evaluated numerically using the forms for $F_{\rm
min}$ discussed in \sec{3}, to give $U_{J,2}=0.61506701$.
The resulting
estimate for the ratio $E^{(2)}/D^2$ is 36.62. This figure is
unreasonably high. Estimates of the mean area\refto{AREA} lead to
$(E^{(1)}/D)^2\approx6.39$, and we would not expect this to be
significantly less than $E^{(2)}/D^2$. Moreover, since the maximal area
of $\pi R^2$ is reached for a circular loop, we would not expect this
amplitude ratio to exceed $\pi^2$.
In fact, recently Guttmann\refto{GPRIV} has estimated a value
for $E^{(2)}$ for the square lattice
which gives $E^{(2)}/D^2\approx6.72$ a much more
reasonable value.

With hindsight it is not hard to find the reason for the failure of the
2-particle approximation in this case. If one examines the region of
integration which is contributing the dominant part to the integral in
\(6o) one finds that it comes from $|\beta|\sim2-5$, which corresponds
to a center of mass energy between $7.5M$ and $150M$, which is far above
the 4-particle threshold at $4M$. Thus, under these circumstances, there
is no justification for ignoring the 4-particle (and perhaps higher)
intermediate states. The fact that the 2-particle contribution
overestimates the result is not surprising in the light of our result
for
$U_0$, where we showed that higher intermediate states can give negative
contributions in the $n\to0$ limit. There appear to be two reasons for
this failure. First, the contribution near the 2-particle threshold is
suppressed by a factor $(1-\cosh2\beta)$, whose origin may be traced to
current conservation. Second, the $\langle J_\mu(r)J_\nu(0)\rangle$
correlation function behaves like $r^{-2}$ in the ultraviolet limit, and
is therefore much more singular than the spin-spin or energy-energy
correlation functions. Thus it is much harder to approximate this
behaviour keeping only the two-particle state, and the
ultraviolet region will give a much more important contribution to the
second moment.

\head{7. Radius of gyration of self-avoiding walks.}

In the previous sections, we have shown how various amplitudes
of self-avoiding walks and loops are related to moments of two-point
correlation functions in the ${\rm O}(n)$ model, which may be estimated
using the form factor approach. In this section we consider another such
quantity, which is, however, related to a higher correlation function.
The
evaluation of this therefore provides a further test of the form factor
method.

Consider and $N$-step self-avoiding walk from the origin to the point
$r$. If $r_i$ labels a site visited by the walk, then the squared radius
of gyration is given by
$$
2N^2R^2_g=\sum_{i,j}(r_i-r_j)^2
\quad.\eqno(7a)
$$
Following the same line of argument as in \sec{2}, if $c_N$ the total
number of $N$-step walks, the generating
function $\sum_Nc_NN^2\langle R^2_g\rangle_Nx^N$
is proportional to the moment
$$
\int(r_1-r_2)^2\langle s(0) {\cal E}(r_1) {\cal E}(r_2)s(r)\rangle\,
d^2r_1d^2r_2d^2r
\quad.\eqno(7b)
$$
The calculation of this four-point function using the form factors would
be very cumbersome: depending on the time-ordering of the points,
different kinds of intermediate states would arise, and, in addition, it
would be necessary to know at least the three-particle form factors of
the spin operator $s$, which we have avoided so far in this paper.

However, the radius of gyration is in fact related to other measures
by the Cardy-Saleur formula\refto{CS} (as corrected by Caracciolo
\etal\refto{Carac}), which is a consequence of the $c$-theorem:
$$
\frac{246}{91}\langle R^2_g\rangle_N -2\langle R^2_m\rangle_N
+\frac12\,\langle R^2_e\rangle_N=0
\quad,\eqno(7c)
$$
where $\langle R^2_m\rangle_N$ is the mean square distance of a monomer
(a site visited by the walk) from one end of the walk, and, as in
\sec{2}, $\langle R^2_e\rangle_N$ is the mean square end-to-end
distance. Thus, if we can compute one universal ratio of these measures,
we may find the other. In fact, $\langle R^2_m\rangle_N$ is easier,
because it involves calculating only a three-point amplitude.

In fact, let us consider the integral
$$
J=\int (r_1+r_2)^2\langle {\cal E}(0)s(r_1)s(r_2)\rangle d^2r_1d^2r_2
\quad.\eqno(7d)
$$
It will become clear below why we choose this particular moment.
Since $(r_1+r_2)^2=2r_1^2+2r_2^2-(r_1-r_2)^2$, this is related to
generating functions in the following way:
$$
J=4\sum_Nc_NN\langle R^2_m\rangle_Nx^{N-1}-
\sum_Nc_NN\langle R^2_e\rangle_Nx^{N-1}
\quad.\eqno(7e)
$$

Let us now see how to evaluate $J$ using the form factor approach. In
general we have to sum over all possible time orderings of
the points $0$, $r_1$ and $r_2$, with a different expression in terms of
the form factors in each region. In addition, the regions where
$t_1$ and $t_2$ are on the same side of the origin will involve, even in
the simplest approximation, knowledge of the form factors
$\langle \beta_1\beta_2|s|\beta_3\rangle$. We may avoid this by the following
trick. Let $r_1=R+\rho$ and $r_2=R-\rho$. If we imagine
doing the $\rho$ integral first, the result will depend only on
$|R|$, so we may choose the time component of $R$ to be zero. This means that
the time ordering will always be $s{\cal E}s$ and not ${\cal E}ss$
or $ss{\cal E}$. Let us therefore write $\rho=(t,x)$ and assume that the
intermediate states are saturated by their lowest contribution, which, in
this time ordering, is only the 1-particle state. Then
$$
\eqalign{
J \approx 4\int 4 R^2\langle {\cal E}(0) s(R-\rho) s(R+\rho)\rangle
d^2R d^2\,\rho & \cr
=64\pi\int_0^\infty R^3dR\int_{-\infty}^\infty dx\int_0^\infty dt
\int{d\beta_1\over2\pi} & \int{d\beta_2\over2\pi}
\langle0|s|\beta_1\rangle e^{-iM(R-x)\sinh\beta_1-Mt\cosh\beta_1}\cr
&\cdot\langle\beta_1|{\cal E}|\beta_2\rangle
e^{iM(R+x)\sinh\beta_2-Mt\cosh\beta_2}\langle\beta_2|s|0\rangle\quad.\cr}
\eqno(7f)
$$
In writing this we have suppressed the colour indices, which are all
set equal to a fixed value by the insertion of the spin operator,
rather than being summed over.
The integrals over $x$ and $t$ are straightforward (it is for this
reason that we choose the particular moment defined by $J$.)
The form factors of the energy operator $\cal E$ are simply proportional
to those of the trace of the stress tensor, $\Theta$, given by
\(FFtrace), so that
$$
\langle\beta|{\cal E}|-\beta\rangle\,=\,
i\,M^2\nu(x_c-x)^{-1} F_{min}(i\pi+2\beta)
\quad.\eqno(7g)
$$
This gives
$$
J=32\pi\nu(x_c-x)^{-1}|\langle0|s|1\rangle|^2
\int_0^\infty R^3dR\int{d\beta\over2\pi}
{e^{-2iMR\sinh\beta}\over
\cosh^2\beta}\,iF_{min}(i\pi+2\beta)
\quad.\eqno(7h)
$$
Before performing the integral over $R$ it is advantageous to deform
the contour so that $\beta\to\beta-i\pi/2$, with the new $\beta$
contour lying just above the real axis. This is possible because of the
analyticity properties of $F_{\rm min}$. The $R$ integral is then
straightforward, and we find
$$
J\approx6\nu(x_c-x)^{-1}M^{-4}|\langle0|s|1\rangle|^2\,I
\quad,\eqno()
$$
where
$$
I=\int_C{iF_{\rm min}(2\beta)\over\cosh^4\beta\sinh^2\beta}d\beta
\quad.\eqno(7i)
$$
The contour runs just above the pole at $\beta=0$. The contribution from
this small semicircle gives $\pi\,\Xi(0)$, where the number $\Xi(0)$ is
evaluated exactly in the Appendix. The remaining principal value
integral may be evaluated numerically using the form for $F_{\rm min}$
given there. The result is that $I\approx\,\pi\,\Xi(0) - 0.456\approx 3.432$.

Now, in the same approximation,
$$\eqalign{
\sum_Nc_N\langle R^2_e\rangle_Nx^N=&\int r^2\langle s(r)s(0)\rangle
d^2r\cr
\approx&|\langle0|s|1\rangle|^2\,3! \int{d\beta\over(M\cosh\beta)^4}
=8M^{-4}|\langle0|s|1\rangle|^2\quad,\cr}
\eqno()
$$
so, for consistency with \(y),
we should take (up to a non-universal normalization which will cancel)
$$
|\langle0|s|1\rangle|^2=M^{\eta}
\quad.\eqno()
$$
Thus, in the one-particle approximation,
$$
\sum_Nc_N\langle
R^2_e\rangle_Nx^N=8\xi_0^{-4+\eta}(1-x/x_c)^{-\gamma-2\nu}
\quad,\eqno()
$$
so that
$$
c_N\langle R^2_e\rangle_N\sim 8\xi_0^{-4+\eta}{N^{\gamma+2\nu-1}\over
\Gamma(\gamma+2\nu)}\mu^N
\quad.\eqno()
$$

Now, using the same normalization, the desired combination of
generating functions \(7e) is
$$
4\sum_Nc_NN\langle R^2_m\rangle_Nx^N-
\sum_Nc_NN\langle R^2_e\rangle_Nx^N=
6\nu I \xi_0^{-4+\eta}(1-x/x_c)^{-\gamma-2\nu-1}
\quad,\eqno(7m)
$$
so that
$$
c_NN(4\langle R^2_m\rangle_N-\langle R^2_e\rangle_N)
\sim 6\nu I \xi_0^{-4+\eta}{N^{\gamma+2\nu+1}\over\Gamma(\gamma+2\nu+1)}\mu^N
\quad,\eqno(7n)
$$
and finally
$$
4{\langle R^2_m\rangle_N\over\langle R^2_e\rangle_N}-1
={6\nu I\over 8}{\Gamma(\gamma+2\nu)\over\Gamma(\gamma+2\nu+1)}
={6\nu I\over 8(\gamma+2\nu)}
\quad.\eqno(7o)
$$
Substituting our estimate for $I$, we then find $\langle
R^2_m\rangle_N/\langle R^2_e\rangle_N\approx 0.420$, which
corresponds to
$\langle
R^2_g\rangle_N\langle R^2_e\rangle_N\approx0.126$. The best
numerical estimates for these ratios, from Monte Carlo simulations,
are $0.43962\pm0.00033$ and $0.14029\pm0.00012$\refto{Carac}.

The approximately 5\% error in the approximation of the first ratio is what
might be expected from the earlier results of this paper. The particular
moment chosen in \(7d) does not damp the short-distance behaviour of the
correlation function as
$r_1$ or $r_2$ become small. On the other hand, the singularities in
this correlation function are less severe than those in the
energy-energy correlation function, and in that case, an estimate of
$U_0$ led to an error of less than 10\%. Unfortunately, the error in
our estimate gets magnified when we use the Cardy-Saleur formula \(7c)
to estimate the other ratio.

Clearly it is possible to improve this calculation at the expense of
considerably more analytic work. However, our result does show that
the utility of the
form factor approach is not restricted to the computation of
two-point functions.

\head{8. Discussion.}

In this paper we have attempted to show that the study of
two-dimensional field theory applied to critical behaviour
has advanced to a point where it is now
capable of yielding not only exact values for critical exponents but
also other universal amplitude ratios and scaling functions which
describe the approach to the critical point. We have chosen the
self-avoiding walk problem as a non-trivial example where
precise numerical comparisons may be made with the results of series
expansions.

Our main results are the estimate for the ratio $C/D$, given in
\(4n); the calculation of the scaling function of the
structure factor for closed loops, whose the coefficients
of whose power series expansion are tabulated in Table I, and whose
asymptotic behaviour is given in \(fasy2,buniversal); and the amplitude
ratio for the radius of gyration of open self-avoiding walks, given in
\sec{7}.

The $S$-matrix and form factor approach to correlation functions, while
not exact in the sense that the sum over intermediate states
must be truncated, turns out to be remarkably useful in practice.
In most cases, only a few-particle intermediate states need be retained
to give a precision of one part in $10^3$. This is related to the
softening of the branch cuts by the interactions.
This effect is peculiar to two dimensions and is a consequence of
two-particle unitarity. Consider, for example, a scalar field theory
with a $g\phi^4$ interaction.
Two-particle phase space is the imaginary part of the simple bubble
diagram
$$
\int{d^2k\over (k^2+M^2)((q-k)^2+M^2)}\sim {1\over M\sqrt{q^2+4M^2}}
\quad,\eqno(8a)
$$
However, in the interacting theory this is modified by the iteration
of the bubble diagram so that it becomes
$$
{1\over M\sqrt{q^2+4M^2}}\,{1\over1+g/(M\sqrt{q^2+4M^2}}\sim
1/g
\eqno(8b)
$$
as $q^2\to-4M^2$. Thus, no matter how small $g$, the two-particle branch
cut is always softened. A similar effect is responsible for the behaviour
$S\to-1$ of the two-body $S$-matrix at threshold. The Watson equations
also show that these two results are manifestations of the same
phenomenon. For higher particle states, the action of this effect in
each two-particle channel separately further suppresses their
contribution.

For the self-avoiding walk problem, this effect has a simple physical
interpretation: the repulsive interaction between different parts of the
walk suppresses those configurations where the walk repeatedly winds back on
itself. It is these which correspond to the multi-particle intermediate
states in the form factor approach.

While this method works well in most of the examples considered
in this paper, it fails in others. The degree of success appears to be
directly related to the softness of the short-distance singularities in the
correlation function, which is given by the conformal dimensions of the
operators involved. For example, for the spin-spin correlation function,
which has a singularity of the form $r^{-5/24}$, even the simple
one-particle approximation does very well, even for the zeroth moment.
For the energy-energy correlation function, which behaves like
$r^{-4/3}$, the 2-particle approximation works to a level of 10\%
for the zeroth moment, and to 1 part in $10^3$ for the second.
The TBA and the $c$-theorem give these moments exactly, in
any case. In the three-point function $\langle s{\cal E}s\rangle$, which has
a singularity $r^{-37/48}$, the level of accuracy is about 5\%. However,
for the current-current correlation function, which behaves like
$r^{-2}$, the two-particle approximation fails completely for the low
moments. It would therefore be useful to have another method of
interpolation between the infrared and ultraviolet behaviours of such
correlation functions.

The possibility of computing higher-point correlation functions also
suggests that these methods may give other interesting geometric
information on the sizes and shapes of self-avoiding walks and loops,
for example, their asphericity.

\noindent{\sl Acknowledgements.} The authors wish to thank A.~J.~Guttmann
for communicating his results on the area of self-avoiding loops,
and the Isaac
Newton Institute for Mathematical Sciences, Cambridge, where this
work was commenced, for its hospitality. This work was supported by the
Isaac Newton Institute, the UK Science and Engineering Research Council,
and the US National Science Foundation under Grant PHY 91-16964.

\references

\refis{BPZ} A.A. Belavin, A.M. Polyakov and A.B. Zamolodchikov,
\journal Nucl. Phys., B241, 333, 1984.

\refis{ISZ} C. Itzykson, H. Saleur and J.B. Zuber, {\sl Conformal
Invariance and Applications to Statistical Mechanics}, (World
Scientific, Singapore 1988).

\refis{Lassig} M. Lassig, \journal Phys. Rev. Lett., 67, 3737, 1991.

\refis{KB} B. Berg, M. Karowski and P. Weisz, \journal Phys. Rev.,
D19, 2477, 1979 ; M. Karowski and P. Weisz, \journal Nucl. Phys.,
B139, 445, 1978; M. Karowski, \journal Phys. Rep., 49, 229, 1979.

\refis{book} F.A. Smirnov, {\sl Form Factors in Completely
Integrable Models of Quantum Field Theory}, (World Scientific), 1992.

\refis{Z1} Al.B. Zamolodchikov, \journal Nucl. Phys., B342, 695, 1990.

\refis{Z2} Al.B. Zamolodchikov, \journal Nucl. Phys., B358, 497, 1991;
\journal Nucl. Phys., B366, 122, 1991.

\refis{Isingb} B.M. McCoy and T.T. Wu, {\sl The two dimensional Ising
model}, (Harvard Univ. Press, 1982).

\refis{Ising} T.T. Wu, B.M. McCoy, C.A. Tracy and M. Barouch,
\journal Phys. Rev., B13, 316, 1976 ; B.M. McCoy, C. Tracy and T.T.
Wu, \journal Phys. Rev. Lett., 38, 793, 1977 ; B.M. McCoy and T.T. Wu,
\journal Phys. Rev. Lett., 45, 675, 1980 ; B.M. McCoy, T.T. Wu,
\journal Phys. Rev., D18, 1259, 1978 ; J.H.H. Perk,
\journal Phys. Lett., 79A, 1, 1990 ; B.M. McCoy, J.H.H. Perk
and T.T. Wu, \journal Phys. Rev. Lett., 46, 757, 1981 ; M. Sato, T. Miwa and
M. Jimbo, \journal Proc. Jap. Acad., 53A, 147, 1977.

\refis{CG} J.L. Cardy and A.J. Guttmann, {\sl Universal Amplitude Combinations
for Self-Avoiding Walks, Polygons and Trails}, NI 92016, OUTP 92-548,
UCSBTH-92-50, to appear in {\sl J. Phys. A: Math. Gen.}

\refis{DG} P.G. de Gennes, \journal Phys. Lett., A38, 339, 1972.

\refis{Nienhuis1} B. Nienhuis, \journal Phys. Rev. Lett., 49, 1062, 1982.

\refis{Nienhuis2} B. Nienhuis, \journal J. Stat. Phys., 34, 731, 1984.

\refis{Nijs} M.P.M. den Nijs, \journal J. Phys., A12, 1857, 1979.

\refis{CH} J.L. Cardy and H.W. Hamber, \journal Phys. Rev. Lett.,
45, 499, 1980.

\refis{DF2} Vl.S. Dotsenko and V.A. Fateev, \journal
Nucl. Phys., B251 [FS 13], 691, 1985; \journal Phys. Lett., B154,
291, 1985.

\refis{DF1} Vl.S. Dotsenko and V.A. Fateev,
\journal Nucl. Phys., B240 [FS 12], 312, 1984.

\refis{Singh} V. Singh and B.S. Shastry, \journal Pramana, 25, 519, 1985.

\refis{Zam1} A.B. Zamolodchikov, in \journal Advanced Studies in Pure
Mathematics, 19, 641, 1989.

\refis{ZZ} A.B. Zamolodchikov and Al.B. Zamolodchikov, \journal Ann.Phys.,
120, 253, 1979.

\refis{Zpoly} A.B. Zamolodchikov, \journal Mod. Phys. Lett., A6,
1807, 1991.

\refis{Smirnovpoly} F.A. Smirnov, \journal Phys. Lett., B275, 109, 1992.

\refis{FF1} F.A. Smirnov, \journal J. Phys., 18, L873, 1984 ;
\journal J. Phys., 19, L575, 1986 ; A.N. Kirillov and F.A. Smirnov,
\journal Phys. Lett., B198, 506, 1987 ; \journal Int. J. Mod. Phys., A3,
731, 1988 ; F.A. Smirnov, \journal Int. J. Mod. Phys., A4, 4231, 1989 ;
\journal Nucl. Phys., B337, 156, 1990.

\refis{FF2} F.A. Smirnov, \journal Comm. Math. Phys., 132, 415, 1990.

\refis{YL} Al. B. Zamolodchikov, \journal Nucl. Phys., B348, 619,
1991.

\refis{CMform} J.L. Cardy and G. Mussardo, \journal Nucl. Phys., B340, 387,
1990; V.P. Yurov and Al.B. Zamolodchikov, \journal Int. J. Mod. Phys.,
A6, 3419, 1991.

\refis{FMS} A. Fring, G. Mussardo and P. Simonetti, \journal
Nucl. Phys., B340, 387, 1993 ; {\sl Form Factors of the Elementary Field
in the Bullough-Dodd Model}, to appear in {\sl Phys. Lett. B}.

\refis{KM} A. Koubek and G. Mussardo, {\sl On the Operator Content of the
Sinh-Gordon Model},
ISAS/EP/93/42, {\sl Phys. Lett. B}, to appear.

\refis{cth} A.B. Zamolodchikov, \journal JETP Lett., 43, 730, 1986;
J.L. Cardy, \journal Phys. Rev. Lett., 60, 2709, 1988.

\refis{CJP} J.L. Cardy, \journal J. Phys. A: Math. Gen., 21, L797, 1988.

\refis{C} A. J. Guttmann and J. Wang, \journal J. Phys. A: Math. Gen.,
24, 3107, 1991.

\refis{Privman} V. Privman and J. Rudnick, \journal J. Phys. A: Math. Gen,
18, L781, 1985.

\refis{Miller} J. Miller, \journal J. Stat. Phys., 63, 89, 1991.

\refis{AREA} S. Leibler, R. R. P. Singh and M. E. Fisher, \journal
Phys. Rev. Lett., 59, 1989, 1987; I. G. Enting and A. J. Guttmann,
\journal J. Stat. Phys., 58, 475, 1990; M. E. Fisher, A. J. Guttmann
and S. G. Whittington, \journal J. Phys. A: Math. Gen., 24, 3095, 1991.

\refis{FSZ} P. Fendley, H. Saleur and Al.B. Zamolodchikov,
{\sl Massless Flow I: the Sine-Gordon and ${\rm O}(n)$ models}, BUHEP-93-5,
USC-93/003,LMP-93-07.

\refis{Cardy} J.L. Cardy, \journal Nucl. Phys., B270 [FS 16], 186, 1986.

\refis{GPRIV} A. J. Guttmann, private communication.

\refis{CS} J.L. Cardy and H. Saleur, \journal J. Phys., A22, L601, 1989.

\refis{Carac} S. Caracciolo, A. Pelissetto and A.D. Sokal, \journal
J.Phys., A23, L969, 1990.


\endreferences

\head{Appendix}

In this appendix we discuss some useful formulas for the $F_{\rm min}(\beta)$
in the limit $n\rightarrow 0$. For large values of $\beta$,
$F_{\rm min}(\beta)$ behaves as
$$
F_{\rm min}(\beta) \,\sim\, e^{\frac{3}{8}\beta} \,\,\, .
\eqno(asyFmin)
$$
Its analytic structure may be read off from its infinite product representation
$$
F_{\min}(\beta)\,=\,\sinh\frac{\beta}{2} \,\,\Xi(\beta)
\eqno(frepr)
$$
where
$$
\Xi(\beta)\,=\,\prod_{l=0}^{\infty}
\left|
\frac{\Gamma(l+1+\frac{i\hat\beta}{2\pi})\,\Gamma(l+2+\frac{i\hat\beta}{2\pi})}
{\Gamma^2\left(l+\frac{3}{2}+\frac{i\hat\beta}{2\pi}\right)}
\frac{\Gamma^2\left(l+\frac{3}{2}\right)}{\Gamma(l+1)\,\Gamma(l+2)}
\right|^{2(l+1)}
\eqno(infinite)
$$
and $\hat\beta=i\pi-\beta$. An interesting quantity is the value of
$\Xi(\beta)$ at the origin
$$
\Xi(0)\,=\,\exp\left(\frac{7}{4\pi^2}\zeta(3)\right)\,=\,1.23756
\eqno(xi0)
$$
(where $\zeta(3)$ is the Riemann function).

For numerical calculations, a useful formula is given by a mixed representation
of $\Xi(\beta)$
$$
\eqalign{
\Xi(\beta) \,=\, &
\prod_{l=0}^{N-1}
\left[\frac{\left(1+\left(\frac{\hat\beta/2 \pi}{l+\frac{3}{2}}\right)^2
\right)^2}
{\left(1+\left(\frac{\hat\beta/2 \pi}{l+1}\right)^2\right)
\left(1+\left(\frac{\hat\beta/2 \pi}{l+2}\right)^2\right)}
\right]^{\frac{(l+1)(l+2)}{2}} \cr
&\times \,
\exp\left[-\int_0^{\infty}
\frac{dx}{x}\frac{\sinh\frac{\pi x}{2}\,q(N,x)\,e^{-2N\pi x}}
{\sinh^2\pi x \,\cosh\frac{\pi x}{2}}\,
\sin^2\frac{\hat\beta x}{2}
\right]\quad, \cr}
\eqno(improved1)
$$
where
$$
q(N,x)\,=\,\frac{(N+1)(N+2)}{2} -N (N+2)\,e^{-2\pi x}+
\frac{(N(N+1)}{2}\,e^{-4\pi x}
\eqno(improved2)
$$
In fact, the rate of convergence of the integral may be improved substantially
by increasing the value of $N$.

\vfill\eject
\centerline{\bf Table I}
\vskip 1cm
\input tables
\begintable
$p$    | ${\cal I}_{2p}$                     | $B_{2p}$ \crthick
$0$    | $2.28627$                           | $1$ \cr
$1$    | $3.53980\,\times 10^{-1}$           | $-0.5$ \cr
$2$    | $1.65158\,\times 10^{-1}$           | $1.17363\,\times 10^{-1}$\cr
$3$    | $1.00385\,\times 10^{-1}$           | $-1.65011\,\times 10^{-2}$\cr
$4$    | $6.92166\,\times 10^{-2}$           | $1.58450\,\times 10^{-3}$\cr
$5$    | $5.14233\,\times 10^{-2}$           | $-1.12325\,\times 10^{-4}$\cr
$6$    | $4.01424\,\times 10^{-2}$           | $6.18847\,\times 10^{-6}$\cr
$7$    | $3.24617\,\times 10^{-2}$           | $-2.74819\,\times 10^{-7}$\cr
$8$    | $2.69541\,\times 10^{-2}$           | $1.01085\,\times 10^{-8}$\cr
$9$    | $2.28462\,\times 10^{-2}$           | $-3.14532\,\times 10^{-10}$\cr
$10$   | $1.96857\,\times 10^{-2}$           | $8.41962\,\times 10^{-12}$\cr
$11$   | $1.71925\,\times 10^{-2}$           | $-1.96581\,\times 10^{-13}$\cr
$12$   | $1.51847\,\times 10^{-2}$           | $4.04936\,\times 10^{-15}$\cr
$13$   | $1.35396\,\times 10^{-2}$           | $-7.43064\,\times 10^{-17}$\cr
$14$   | $1.21715\,\times 10^{-2}$           | $1.22477\,\times 10^{-18}$\cr
$15$   | $1.10193\,\times 10^{-2}$           | $-1.82639\,\times 10^{-20}$\cr
$16$   | $1.00382\,\times 10^{-2}$           | $2.47954\,\times 10^{-22}$\cr
$17$   | $9.19444\,\times 10^{-3}$           | $-3.08162\,\times 10^{-24}$\cr
$18$   | $8.46266\,\times 10^{-3}$           | $3.52340\,\times 10^{-26}$\cr
$19$   | $7.82306\,\times 10^{-3}$           | $-3.72244\,\times 10^{-28}$  \cr
$20$   | $7.26018\,\times 10^{-3}$           | $3.64834\,\times 10^{-30}$
\endtable
\vfill
\tenpoint
\noindent Table I. The values of the first twenty
integrals ${\cal I}_{2p}$ defined
in \(I2p), and the coefficients $B_{2p}$ of the structure factor scaling
function \(final), expanded in powers of its argument $y=qR$.

\endit